\documentclass[11pt,twocolumn,dvipsnames]{aastex62}
\usepackage{xcolor}
\usepackage{lineno}
\usepackage{comment}

\usepackage{xcolor}
\usepackage[colorlinks=true]{hyperref}
\hypersetup{
    colorlinks=true,
    linkcolor=RoyalBlue,
    filecolor=magenta,      
    urlcolor=orange,
    citecolor=BrickRed
}

\shorttitle{High-Energy Neutrinos from Seyfert Galaxies}

\begin{document}

\title{High-Energy Neutrinos from Magnetized Coronae of Active Galactic Nuclei and\\
Prospects for Identification of Seyfert Galaxies and Quasars in Neutrino Telescopes}

\author{Ali Kheirandish}
\thanks{kheirandish@psu.edu}
\affiliation{Department of Physics; Department of Astronomy \& Astrophysics; Center for Multimessenger Astrophysics, Institute for Gravitation and the Cosmos, The Pennsylvania State University, University Park, PA 16802, USA}
\author{Kohta Murase}
\thanks{murase@psu.edu}
\affiliation{Department of Physics; Department of Astronomy \& Astrophysics; Center for Multimessenger Astrophysics, Institute for Gravitation and the Cosmos,
The Pennsylvania State University, University Park, PA 16802, USA\\
\& \\
Center for Gravitational Physics, Yukawa Institute for Theoretical Physics, Kyoto University, Kyoto, Kyoto 606-8502, Japan\\}
\author{Shigeo S. Kimura}
\thanks{shigeo@astr.tohoku.ac.jp}
\affiliation{Frontier Research Institute for Interdisciplinary Sciences, Tohoku University, Sendai 980-8578, Japan\\
\&\\
Astronomical Institute, Tohoku University, Sendai 980-8578, Japan}

\begin{abstract}
Particles may be accelerated in magnetized coronae via magnetic reconnections and/or plasma turbulence, leading to high-energy neutrinos and soft gamma rays. We evaluate the detectability of neutrinos from nearby bright Seyfert galaxies identified by X-ray measurements.  In the disk-corona model, we find that NGC 1068 is the most promising Seyfert galaxy in the Northern sky, where IceCube is the most sensitive, and show prospects for the identification of aggregated neutrino signals from Seyfert galaxies bright in X-rays. Moreover, we demonstrate that nearby Seyfert galaxies are promising targets for the next generation of neutrino telescopes such as KM3NeT and IceCube-Gen2. 
For KM3NeT, Cen A can be the most promising source in the Southern sky if a significant fraction of the observed X-rays come from the corona, and it could be identified in few years of KM3NeT operation. Our results reinforce the idea that hidden cores of supermassive black holes are the dominant sources of the high-energy neutrino emission and underlines the necessity of better sensitivity to medium-energy ranges in future neutrino detectors for identifying the origin of high-energy cosmic neutrinos.
\end{abstract}


\keywords{astroparticle physics -- galaxies: active -- galaxies: jets -- gamma rays: galaxies -- neutrinos -- radiation mechanisms: non-thermal}

\section{Introduction} \label{sec:intro}
The observation of high-energy cosmic neutrinos in IceCube~\citep{Aartsen:2013bka, Aartsen:2013jdh} has revealed that hadronic interactions play a major role in the non-thermal emission in the high-energy universe. While the magnitude of the measured spectrum at high energies, $\gtrsim100$ TeV, was found to be consistent with some theoretical predictions for astrophysical neutrinos~\citep{Loeb:2006tw,Murase:2008yt,Kotera:2009ms,Murase:2013rfa}, the emergence of an order of magnitude higher flux at medium energies in the 10~TeV range \citep{Aartsen:2014muf, Aartsen:2015knd, Aartsen:2020aqd} pointed towards sources with efficient neutrino emission. 
While the dominant sources of high-energy cosmic neutrinos are yet to be identified, the multimessenger data indicate that their sources should be opaque to high-energy $\gamma$-rays in the GeV-TeV range, i.e., hidden to the $\gamma$-ray telescopes operating at these energies~\citep{Murase:2015xka,Capanema:2020rjj,Capanema:2020oet}. Otherwise, the contribution of the $\gamma$-ray flux leaving transparent sources to the diffuse isotropic $\gamma$-ray background (IGRB) would create an excess that would overshoot the measured diffuse flux by {\em Fermi}-LAT~\citep{Ackermann:2014usa}. { Consequently, scenarios such as galaxy clusters/groups and starburst galaxies as the dominant contributors to the high-energy cosmic neutrino flux below 100 TeV are excluded because they will overwhelm the observed IGRB flux \citep{Senno:2015tra,Murase:2015xka,Bechtol:2015uqb}. On the other hand, the flux above 100 TeV can still be explained by such cosmic-ray (CR) reservoir scenarios \citep{Murase:2013rfa,Xiao:2016rvd,Murase:2016gly,Fang:2017zjf,Liu:2017bjr,Peretti:2019vsj}.
}

Cores of the active galactic nuclei (AGN), which are optically thick for GeV-TeV $\gamma$-rays are one of the best candidates as the source of the high-energy neutrino flux at medium energies. 
The isotropic distribution of the arrival direction of high-energy neutrinos, together with the constraints on the Galactic component of the neutrino flux, points at extragalactic sources {\em hidden} in $\gamma$-rays as the origin of this flux. Moreover, in order to produce this level of flux, efficient neutrino production is required. Such conditions can be met with large column densities of target material and radiation, which form dense environments suppressing the $\gamma$-ray flux at the site of production.
 
In Seyfert galaxies, accretion dynamics and magnetic dissipation will form a magnetized {\em corona} above the disk, see {\em e.g.,} \cite{Miller:1999ix,Merloni:2000gs,Liu:2002ts,Blackman:2009fi,Io:2013gja,Jiang:2014wga,2019ApJ...885..144J}. 
Thanks to the dense environments near the supermassive black holes, the accelerated CRs in the corona interact efficiently with the gas and target photons, producing charged and neutral pions that promptly decay to neutrinos and $\gamma$-rays. The neutrinos will escape the environment while GeV-TeV $\gamma$-rays are cascaded down inside the source.

The disk-corona model for high-energy neutrino emission from the core of AGNs can successfully accommodate the flux of cosmic neutrinos at medium energies in the $10-100$ TeV range \citep{Murase:2019vdl}. In addition, modeling of the high-energy emission from the hidden cores of AGNs presents a robust connection to $\gamma$-ray emission at MeV energies. 

The dominant role of AGN cores to the flux of high-energy neutrinos at medium energies has been reinforced by the IceCube's most recent search for the sources of high-energy neutrinos using 10 years of data \citep{Aartsen:2019fau}. The all-sky (untriggered) and the source list (triggered) time-integrated analysis of the data finds the direction of NGC 1068 as the most significant point in the sky. The excess in the direction of NGC 1068 is found { local pre-trial p-value of $1.8\times 10^{-5}$, yielding a 2.9$\sigma$} after correcting for trials.

NGC 1068, aka Messier 77, is a nearby, bright Seyfert 2 galaxy. The observed infrared luminosity of NGC 1068 is similar to starburst galaxies. It has been considered as a potential CR accelerator and motivated study of neutrino and $\gamma$-ray emission, see {\em, e.g.,} \citet{Yoast-Hull:2013qfa, Murase:2016gly,Liu:2017bjr,Lamastra:2019zss}. The best-fit neutrino flux measured by IceCube, however, exceeds the observed $\gamma$-ray emission by {\em Fermi}-LAT \citep{Ackermann_2012}. Therefore, the models that are built upon the measured $\gamma$-ray emission in {\em Fermi} cannot accommodate the reported neutrino flux. In the meantime, neutrino flux predictions from starburst activity are directly correlated with the {\em Fermi} $\gamma$-rays and face a similar obstacle.

In this paper, we present the expected neutrino emission from the brightest Seyfert galaxies in the disk-corona model by incorporating the measured intrinsic X-ray luminosities. We estimate the parameters relevant for emission from the magnetized coronae and establish the parameter space compatible with the reported neutrino spectrum of NGC 1068 and the medium-energy excess in the cascade data.  We further employ the parameters that can accommodate the emission from NGC 1068 for the brightest Seyfert galaxies and study their likelihood of observation, individually and when stacked together. 

While providing a consistent level of neutrino flux, our modeling finds NGC 1068 as the most promising source in the Northern hemisphere, where IceCube is most sensitive. While the majority of the bright sources reside in the Southern sky, we show that the stacking analysis has a good chance of identifying neutrino emission from this class of sources. In addition, the significance of the search for neutrino emission from such analysis can inform the principal mechanism for the high-energy emission.

We use the predicted neutrino emission to examine the prospects for observation of the nearby, bright Seyfert galaxies in future neutrino telescopes, KM3NeT and the next generation of IceCube at the South Pole: IceCube-Gen2.

While indicating that Seyfert galaxies are the dominant contributors to the high magnitude flux of cosmic neutrinos at energies below 100 TeV, our prediction provides a testable scenario in current and future neutrino telescopes. The subset of bright Seyfert galaxies is likely to be observed with the continuous operation of the IceCube detector and commissioning of the neutrino telescopes in the Northern hemisphere. 

In the next section, we discuss details of acceleration mechanism and neutrino production in the magnetized corona model. Then, in Sec.~\ref{sec:nuseyf} we present the neutrino emission under 3 distinct scenarios from NGC 1068 and the brightest Seyfert galaxies identified by the X-ray surveys. Subsequently, in Sec.~\ref{nextgen}, we will study the prospects for identification of the neutrino emission in current and future neutrino telescopes. Finally, we will discuss the implications of observing neutrinos from these sources and the relation to the total flux in Secs.~\ref{sec:impl}.\\

\section{Magnetized Corona Model}\label{sec:model}
Progress in X-ray observations of AGN led to the establishment of the magnetized coronal paradigm \citep{Haardt:1991tp,Liu:2002ps,2018MNRAS.480.1819R}. This is partly confirmed by recent magnetohydrodynamic (MHD) simulations \citep[e.g.,][]{Io:2013gja,Jiang:2019bxn}. The magnetorotational instability (MRI; \citealt{Balbus:1991ay}) supports that the coronae above the optically thick accretion disk are naturally expected to be hot, magnetized, and turbulent. 
Such a plasma can be collisionless in the sense that the accretion onto a central black hole is faster than their Coulomb collisions time, and ions may be accelerated via stochastic acceleration \citep[e.g.,][]{lyn+14,KTS16a,2018PhRvL.121y5101C,Zhdankin:2018lhq,2019MNRAS.485..163K,Wong:2019dog,Lemoine:2020bsk} and/or magnetic reconnection processes \citep[e.g.,][]{Hoshino:2013pza,hos15,2016ApJ...818L...9G,2018MNRAS.473.4840W,Ball:2018icx}. 
{ Stcochastic acceleration and magnetic reconnection may naturally coexist in highly magnetized plasmas. Although either may become a dominant channel for particle acceleration \citep{Comisso:2019frj,Zhdankin:2018lhq,2021PhPl...28e2905L}, they cannot easily be separated. In this study, for demonstration, we consider particle acceleration and neutrino production under each scenario separately.}

\subsection{Stochastic acceleration scenario}
It is widely believed that turbulent magnetic fields generated by the MRI is responsible for the angular momentum transport in accretion flows, which has been confirmed by MHD simulations \citep{SP01a,MM03a,OM11a,2012MNRAS.426.3241N,2019ApJS..243...26P}. Turbulence can also be generated through magnetic reconnections resulting from magnetic fields amplified by the MRI. CRs in the MHD turbulence randomly change their energies through interactions with MHD waves. If the acceleration in momentum space homogeneously occurs without significant trapping \citep{Lemoine:2020bsk}, the phenomenon can be described by the following diffusion equation in energy space: 
\begin{equation}
  \frac{\partial {\mathcal F}_p}{\partial t} = \frac{1}{\varepsilon_p^2}\frac{\partial}{\partial \varepsilon_p}\left(\varepsilon_p^2D_{\varepsilon_p}\frac{\partial {\mathcal F}_p}{\partial \varepsilon_p} + \frac{\varepsilon_p^3}{t_{\rm cool}}{\mathcal F}_p\right) -\frac{{\mathcal F}_p}{t_{\rm esc}}+\dot {\mathcal F}_{p,\rm inj},\label{eq:FP}  
\end{equation}
where $\mathcal{F}_p$ is the CR distribution function ($dN/d\varepsilon_p=4\pi p^2\mathcal{F}_p/c$), $D_{\varepsilon_p}$ is the diffusion coefficient in energy space, $t_{\rm cool}$ is the cooling timescale, and $t_{\rm esc}$ is the escape timescale. 
The injection function is given by $\dot {\mathcal F}_{p,\rm inj} = {f_{\rm inj}L_X \delta(\varepsilon_p-\varepsilon_{\rm inj})}/[{4\pi {(\varepsilon_{\rm inj}/c)}^3 {\mathcal V}}],$ where $\varepsilon_{\rm inj}$ is the injection energy and $f_{\rm inj}$ is the injection fraction at $\varepsilon_{\rm inj}$, $L_X$ is the X-ray luminosity, $\mathcal{V}$ is the volume of the corona, and $\delta(x)$ is the delta function. {Note that in this scenario, the injection fraction is proportional to the dissipation rate in the corona, and therefore, is proportional to $L_X$.}
The diffusion coefficient is assumed to scale as $D_{\varepsilon_p}\propto \varepsilon_p^q$, where $q$ is the power-law index of the diffusion coefficient in momentum space. 
If the gyro-resonant scattering is dominant, which is not necessarily the case, it corresponds to the spectral index of the turbulence power spectrum. For demonstrative purposes, we use $q=5/3$ motivated by the Kolmogorov turbulence, but $q=3/2$ (that can be expected for isotropic MHD turbulence in the magnetically dominated regime) or $q=2$ \citep[that is a hard-sphere type; see, e.g.,][]{2019MNRAS.485..163K} may be possible. Note that in general the CR spectrum can be a power law if the trappiing is significant \citep{Lemoine:2020bsk}.
As cooling processes, we take into account Bethe-Heitler, $pp$ inelastic collisions, and photomeson ($p\gamma$) production processes. We consider advective (infall to a central black hole) and diffusive escapes. {The maximum energy of the CRs is set by the balance between the acceleration and these losses. For the latter, the principal component is the Bethe-Heitler process due to abundant disk photons, as emphasized in \citet{Murase:2019vdl}. This results in a cutoff feature in the CR spectrum below PeV energies.} In this scenario, we introduce two parameters, the CR pressure to the thermal pressure $P_{\rm CR}/P_{\rm th}$ and turbulent strength $\eta_{\rm tur}^{-1}$. The former determines the normalization of CRs, whereas the latter is related to the maximum energy of CRs. 
Note that we use the thermal pressure at the virial temperature, i.e., $P_{\rm th}=n_p kT_{\rm vir}$, where $n_p$ is the thermal proton density and $T_{\rm vir}$ is the virial temperature, and the realistic thermal pressure can be reduced if the ion temperature is lower than $T_{\rm vir}$.  
See Supplemental Material of \cite{Murase:2019vdl} for details \citep[see also Section III-B of][in the context of radiatively inefficient accretion flows]{2019PhRvD.100h3014K}.

\subsection{Magnetic reconnection (fast acceleration) scenario}
Recent Particle-In-Cell (PIC) Simulations revealed that relativistic magnetic reconnections in the ion-electron plasma, where the ion magnetization parameter $\sigma_i=B^2/(8\pi n_pm_p c^2)\gtrsim 1$, can accelerate non-thermal relativistic particles very efficiently, given that the plasma $\beta$, which is defined by $\beta\equiv8\pi n_p kT_p/B^2$, is smaller than $\sim0.01$ \citep{2016ApJ...818L...9G,2018MNRAS.473.4840W,Ball:2018icx}. 
For AGN coronae, $\beta$ can be sufficiently small (e.g., $\beta\lesssim1-3$) but we typically expect that the plasma is semi-relativistic, $\sigma_i=2kT_p/(m_pc^2\beta)<1$ \citep{2018MNRAS.473.4840W}. It is still uncertain whether protons can be accelerated above $\sim m_pc^2$, and further acceleration via turbulence would be necessary \citep{2018PhRvL.121y5101C,Zhdankin:2018lhq,Wong:2019dog,Lemoine:2020bsk}. Not only the stochastic acceleration but also the first-order Fermi acceleration may operate due to scatterings with reconnection outflows \citep[e.g.,][]{Pisokas:2017zxx,PhysRevLett.108.135003}. In this work, for the purpose of phenomenological studies, we simply assume that the injection spectrum of protons is given by a single power-law with an exponential cutoff, providing the power-law index, $s$, as a parameter. 
{ We assume that the corona can be regarded as electron-proton plasma. X-ray observational data seem consistent with the electron-proton plasma \citep{2018MNRAS.480.1819R}, although better quality data are necessary to clarify it. For electron-positron-proton plasma, the non-thermal proton production efficiency via magnetic reconnections is expected to be lower than that for electron-proton plasma \citep[e.g.,][]{Petropoulou:2019bse}.}

To calculate neutrino emission, we solve the transport equation for the non-thermal protons with single-zone and steady-state assumptions:
\begin{equation}
-\frac{d}{d\varepsilon_p}\left(\frac{\varepsilon_pN_{\varepsilon_p}}{t_{\rm cool}}\right)=\dot{N}_{\varepsilon_p,\rm inj}-\frac{N_{\varepsilon_p}}{t_{\rm esc}},
\end{equation}
where $N_{\varepsilon_p}=dN/d\varepsilon_p$, $\dot{N}_{\varepsilon_p,\rm inj}\propto\varepsilon_p^{-s}$ is the injection term.

We consider the same cooling and escape processes as those in the stochastic acceleration scenario. { The maximum proton energy of the injection spectrum is determined by either the size of the acceleration region or the balance between acceleration and the total losses via cooling and escape processes, as $E_p^{\rm max}={\rm min}[E_p^{\rm rec},E_p^{\rm cool}]$. In general, particle acceleration consists of multi-phase processes including the electric field acceleration and Fermi acceleration~\citep{2021PhPl...28e2905L}. 
In the absence of losses, the former energy can be written as $E_p^{\rm rec}\sim eB l_{\rm rec}$, where $l_{\rm rec}$ is the reconnection region thickness that can be smaller than the system size~\citep{PhysRevLett.108.135003,2016MNRAS.463.4331D}.    
The latter is determined by $t_{\rm acc}=t_{\rm cool}$, where $t_{\rm acc}$ is the proton acceleration time and $t_{\rm cool}$ is the proton cooling time. 
The acceleration time depends on details of the magnetic reconnection process. For example, particle bounces may occur either in converging magnetic fluxes with reconnection velocity~\citep{2005A&A...441..845D,2016MNRAS.463.4331D,Medina-Torrejon:2020fae} and/or by outflows with Alfv\'en velocity~\citep{2006Natur.443..553D,PhysRevLett.108.135003}. However, recent simulations have shown that the acceleration is efficient especially for relativistic reconnections~\citep{HL12}. For demonstrative purposes, we assume $t_{\rm acc}=\eta_{\rm acc}r_L/c$, where $\eta_{\rm acc}$ is the acceleration efficiency parameter and $r_L$ is the Larmor radius. At sufficient high energies the Fermi-like acceleration, and $\eta_{\rm acc}$ can be as small as $\sim3(c/V_{\rm rec})$~\citep{2010MNRAS.408L..46G,Zhang:2021akj}, although details are uncertain in non-relativistic reconnections~\citep[c.f.][]{PhysRevLett.108.135003,2016MNRAS.463.4331D}. In this work, we assume $\eta_{\rm acc}=300$, but we will see that $E_p^{\rm max}=E_p^{\rm rec}$ to better explain the observation of NGC 1068 (see Section~\ref{sec:NGC1068}).} 
The normalization of CRs are provided such that $\int \dot{N}_{\rm inj}\varepsilon_p d\varepsilon_p = \epsilon_{\rm CR} \dot{M}c^2$, where $\epsilon_{\rm CR}$ is the energy fraction carried by CRs and $\dot{M}$ is the mass accretion rate in the accretion disk. { Only the fraction of the accretion energy is spent to accelerate CRs. Therefore, we use the total mass accretion rate onto the BH to deduce the CR normalization. The total mass accretion rate is proportional to the bolometric luminosity, which is estimated from intrinsic X-ray luminosity}. Note that the corona is connected to the accretion disk through magnetic fields, and a significant fraction of the accretion energy may be dissipated in the disk-corona interface. 

 The previous works focused on the stochastic acceleration \citep{Murase:2019vdl} or shock acceleration mechanism (see below), and the magnetic reconnection scenario was not considered in detail. In this work, we show that the diffuse neutrino flux can be explained by the magnetic reconnection scenario in the AGN corona model. In Sec.~\ref{sec:summary} we discuss the possibility of explaining the diffuse neutrino flux at medium energies with this scenario and comment on the feasibility of identifying this scenario in the near future.

\subsection{Remarks on other acceleration scenarios}
This work focuses on the standard magnetized corona scenario based on recent global MHD simulations, which does not involve accretion shocks. Alternatively, accretion shocks have been discussed as a CR production site in the accretion flows~\citep{brs90,Stecker:1991vm,2019ApJ...880...40I}. Originally, this scenario was proposed to explain X-ray observations of AGNs by CR-induced electromagnetic cascades~\citep{1981MNRAS.194....3B,1986ApJ...305...45Z}. However, cutoff and softening features were discovered in bright AGNs in 1990s~\citep{1993ApJ...407L..61M,2000ApJ...542..703Z}, which ruled out the hadron-induced electromagnetic cascade scenario for the observed X-rays. The accretion shock could still be viable to be a neutrino production site as long as the CR-induced cascade flux is below the X-ray and gamma-ray data~\citep{Stecker:2005hn,Stecker:2013fxa,2019ApJ...880...40I}. However, there are several problems in this scenario. 
First, the accretion shocks are not seen in any global MHD simulations dedicated for accretion flows~\citep[e.g.,][]{2012MNRAS.426.3241N,ktt14,Jiang:2019bxn}. Although there are solutions with an accretion shock in the one-dimensional hydrodynamic equation system with a steady state assumption~\citep{bdl11,Chattopadhyay:2016kcz}, such solutions are not realized in these dedicated simulations. 
The second point is the angular momentum of the accreting matter. For example, \citet{2019ApJ...880...40I} explicitly assumed the free-fall spherical accretion (see their Equation~2). However,. we expect a large specific angular momentum at an outer scale, which prevents the matter from freely falling to a vicinity of the supermassive black hole. An anomalously efficient angular momentum transport mechanism is necessary to form an accretion shock, which contradicts with the free-fall assumption made by \cite{2019ApJ...880...40I,Inoue:2019yfs}. Lastly, the accretion rate through the accretion shocks is assumed to be extremely high to explain the IceCube data~\citep{2019ApJ...880...40I}. It is comparable to or even higher than the accretion rate expected in the geometrically thin, optically thick disk, which requires two accretion components with totally different angular momentum distributions.
Note that $P_{\rm CR}<0.5P_{\rm th}$ is required in this scenario, otherwise hot coronae cannot be maintained.

We also note that the Bethe-Heitler process, which was often ignored, is relevant. The $10-100$~TeV neutrino flux from the photomeson production with coronal X-rays is suppressed by Bethe-Heitler interactions with disk photons~\citep{Murase:2019vdl}. On the contrary, given that shock acceleration leads to higher maximum energies, PeV neutrinos are mainly produced via the photomeson production with disk photons~\citep{Stecker:1991vm}. 

Also, the realistic coronal structure is inhomogeneous. 
As shown in \cite{Gutierrez:2021vnk}, radio emission is more likely to come from outer coronal regions. Such a multi-zone situation is commonly invoked to explain radio emission from blazars and radio galaxies because it is known that radio emission cannot be explained by single-zone models.  

Note that \cite{2015JETP..120..541K} assumed electric acceleration in a gap formed in the magnetosphere. The gap formation is possible only in low-luminosity AGNs such as M87, but it is likely to be screened in the case of Seyferts and quasars because the accretion rate is so high that the plasma satisfies the quasi-neutrality condition \citep{Levinson:2010fc}. 

\section{Neutrino emission from bright Seyfert Galaxies}\label{sec:nuseyf}
The high-energy neutrino flux modeled in this study incorporates both hadronuclear ($pp$) and photohadronic ($p\gamma$) interactions. The inclusion of both the processes in the production of high-energy neutrinos, together with the cooling processes in the source, results in features in the neutrino flux from Seyfert galaxies. Thus, the spectrum of high-energy neutrinos would deviate from simplistic single power laws.  

The {\em stochastic acceleration} scenario for the production of high-energy neutrinos may yield a spectrum that peaks at $\sim 10$ TeV energies and steeply falls around $\sim$ 100 TeV. This feature deviates sharply from the single power-law assumption for the spectrum incorporated in the search for neutrino sources. Meanwhile, the other scenario considered here, the {\em magnetic reconnection} scenario, can yield a neutrino spectrum that mimics a power-law with a cutoff at high energies. 
Neutrino production in both scenarios may also receive a significant contribution from the photomeson production process. This often leads to a bumpy feature before the cutoff in the spectrum.

In this section, we first estimate the neutrino flux from NGC 1068 and adopt parameters in each acceleration scenario that can describe the reported neutrino spectrum from the source while maintaining other constraints on the parameters in a physical range. The main parameter in our modeling that constrains the neutrino spectrum is the ratio of the CR to thermal pressures. After finding the parameters for the neutrino flux that can accommodate the neutrino flux from NGC 1068, we estimate the neutrino flux for the brightest Seyfert galaxies, similar to NGC 1068.  
Using the BAT AGN Spectroscopic Survey (BASS), we select the ten sources that are classified as Seyfert galaxies and pose the highest X-ray flux. We incorporate the X-ray luminosity measured in 2-10 keV band. This range suits our selection criteria as we are considering the brightest X-ray sources whose luminosities are $L_X \gtrsim 10^{42}$ erg/s.  
Less bright sources such as Sgr A*, with $L_X < 10^{39} \rm erg/s$ are better characterized by measurements at lower-energy bands. 
{ Throughout this work, we estimate the distance of the Seyfert galaxies based on their redshift provided by BASS. To evaluate the luminosity distance, we employ the following cosmological parameters: $H_0=70\, \rm km\,s^{-1}\, Mpc^{-1}$, $\Omega_m=0.3$, $\Omega_\Lambda=0.7$}.

\subsection{Neutrino emission from NGC 1068}\label{sec:NGC1068}
NGC 1068 is a Seyfert 2 galaxy at $z=0.00303$ \citep{1988ngc..book.....T}, with a heavily obscured nucleus. It is considered as one of the best studied AGNs, thanks to its role in the AGN unification scenarios, which was initially proposed for explaining the broad optical lines in its polarized light. The column density of NGC 1068 is $\gtrsim 10^{25}\rm~cm^{-2}$, implying a Compton thick environment. 
The high density environment at the core of NGC 1068 and the high level of mass accretion rate provide favorable environments for efficient production of high-energy neutrinos.

In order to estimate the neutrino spectrum in the AGN corona model for NGC 1068, we use the intrinsic X-ray luminosity, $L_X$, suggested by {\em Swift}-BAT AGN Spectroscopic Survey (BASS) \citep{Ricci:2017dhj}. The BASS catalog provides X-ray fluxes and luminosities in a wide range, covering both soft (below 2 keV) and hard (above 10 keV) X-rays. Here, we use measured spectrum and luminosity in the 2-10 keV band, which provides a medium range. This choice suits the disk-corona modeling, as the abundance of data in this energy band enables us to use empirical correlation.

The intrinsic X-ray luminosity, $L_{\rm X}$ measured from the direction of NGC 1068 by BAT is $\sim10^{43} ~\rm erg/s$ for the 2-10 keV band, although it could be larger depending on estimates of the column density. We use this luminosity to model the neutrino spectrum from NGC 1068 in three different scenarios. 

Assuming a single power-law spectrum, the IceCube Collaboration has reported a best-fit flux of $3\times 10^{-11}\, \rm TeV^{-1} cm^{-2}s^{-1}$ at 1 TeV with an index of $\simeq$ 3.2 \citep{Aartsen:2019fau}. As a criterion, we adopt parameters that maintain the flux within the best-fit reported flux and its uncertainties reported by the IceCube Collaboration.

First, we model neutrino production assuming the {\em stochastic acceleration} scenario. As mentioned earlier, in this scenario, the neutrino spectrum has a more complicated shape than a single power-law. Accommodating the IceCube flux at TeV energies requires a relatively high normalization while the spectrum has to cutoff fast enough that the spectrum drops around 100 TeV. Such conditions would result in a high level of CR pressure in the corona model.

In order to maintain realistic scenarios, we restrict ourselves to the range of parameters for which the ratio of the CR pressure ($P_{\rm CR}$) to the thermal pressure ($P_{\rm th}$) is bound to less than 0.5. In this limit, the non-thermal energy is equal to half of the gravitational binding energy at the coronal radius without leaving room for thermal particles. Although the coronal plasma may be heated more through magnetic fields connected to the inner disk, we assume 0.5 as the maximal case in this work, and the neutrino spectrum peaks at $\sim$ 5 TeV and falls sharply around 20 TeV. We refer to this scenario as ``High CR pressure"

\begin{figure}[t]
    \centering
    \includegraphics[width=\columnwidth]{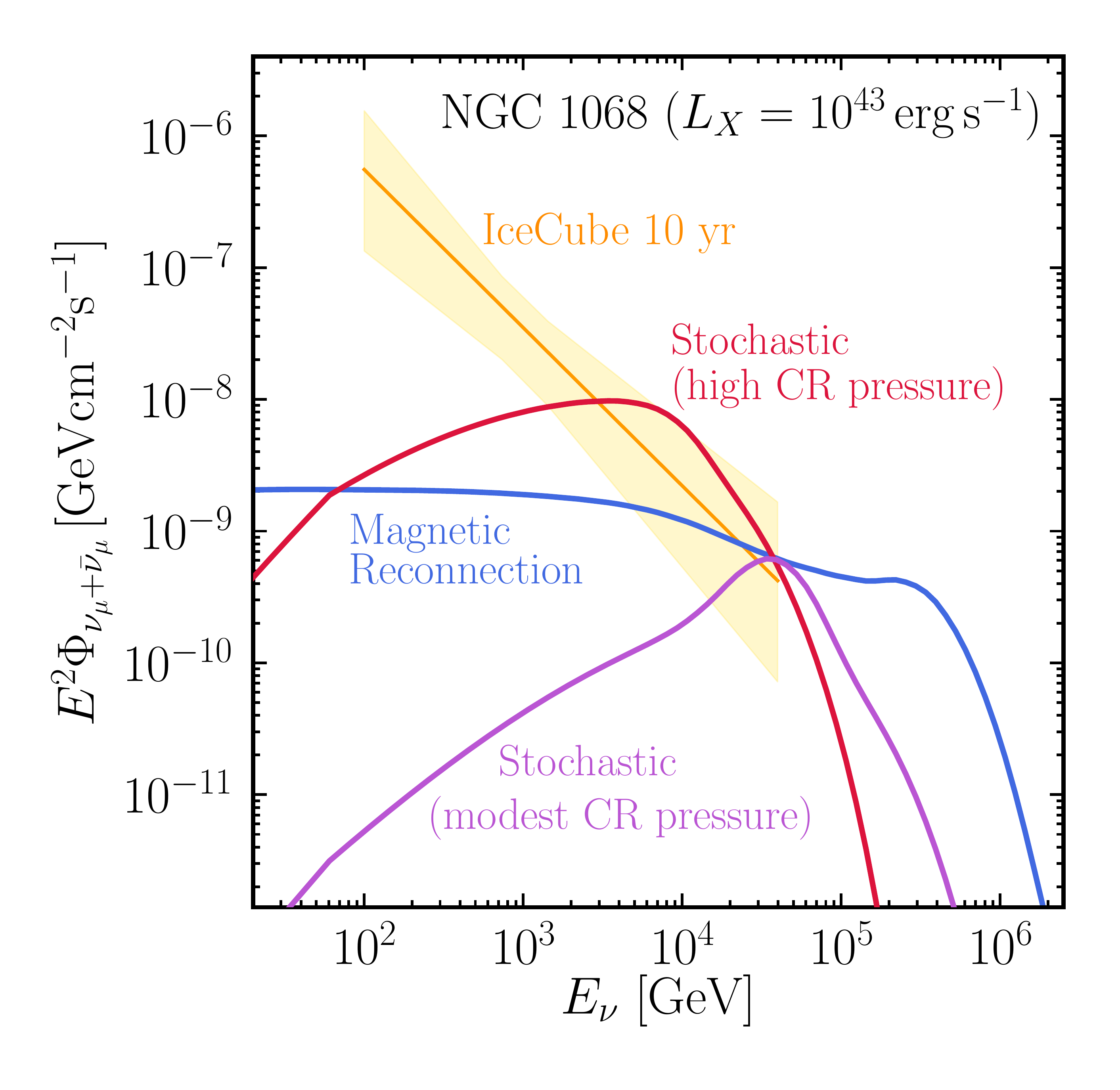}
    \caption{Modeled neutrino spectrum for NGC 1068 compared to the best fit flux (yellow band) reported by the IceCube Collaboration 10 yr point source study \citep{Aartsen:2019fau}. The red line shows the expected flux in the stochastic acceleration scenario matching IceCube's best fit at TeVs. The purple line depicts the flux that would give the medium-energy neutrino flux, compatible with the total neutrino flux reported in the cascade analysis \citep{Aartsen:2020aqd}. The blue line presents the flux expected for the magnetic reconnection scenario.}
    \label{fig:ngc1068nuflux}
\end{figure}

We consider the second scenario for neutrino emission from NGC 1068 assuming coronal emission from stochastically accelerated particles, where instead of matching the flux at TeVs, we match the diffuse neutrino flux at tens of TeV, motivated by the medium-energy excess in the neutrino spectrum. In this case, as shown previously \citep{Murase:2019vdl} we adopt parameters that can explain the high-energy neutrino flux excess observed at medium energies \citep{Aartsen:2020aqd}. In this case, the $P_{\rm CR}/P_{\rm th}$ is set to $\simeq 0.01$. Here, the neutrino spectrum peaks at $\sim 40$ TeV, which corresponds to a lower level of neutrino flux compared to the previous scenario. We refer to this case as ``Modest CR pressure" hereafter.

These results are compatible with the spectra presented previously by \citet{Murase:2019vdl} where the CR pressure considered to explain the medium-energy neutrino flux and NGC 1068 are found at the level of $\sim$ 1 and $\sim$ 30 percent of the thermal pressure, respectively. Here, we allow the pressure ratio to be as high as 50\% to explain the soft spectrum reported for NGC 1068 by the IceCube Collaboration \citep{Aartsen:2019fau}.
Note that, in principle, both the High CR pressure and Modest CR pressure cases can be viable within the same stochastic acceleration scenario. For example, the Modest CR pressure may be realized in average AGN, whereas some sources such as NGC 1068 may have a large CR pressure.  

\begin{figure}[t]
    \centering
    \includegraphics[width=\columnwidth]{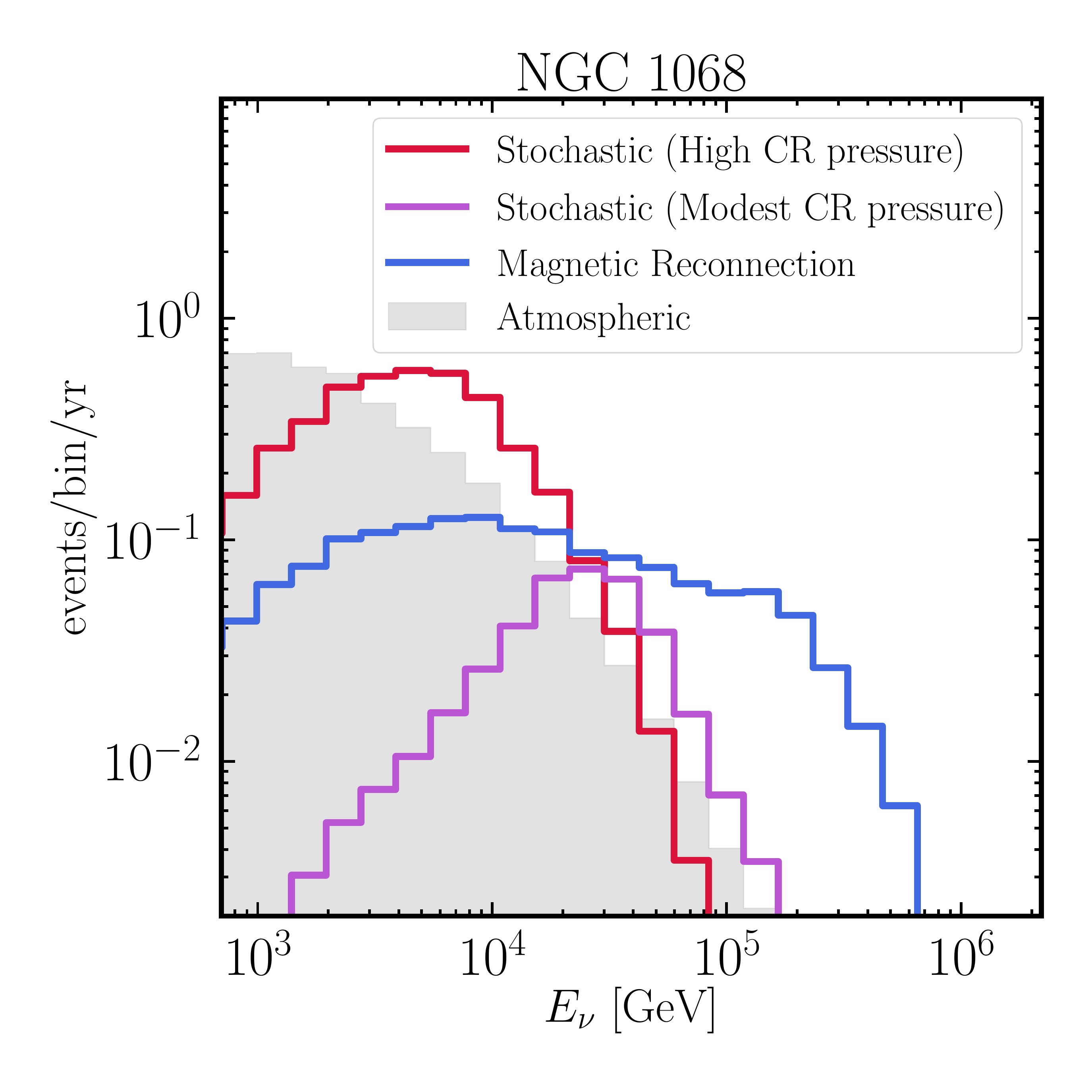}
    \caption{Expected event distribution per energy and year in IceCube is shown for the stochastic and reconnection scenarios presented for NGC 1068 in Figure \ref{fig:ngc1068nuflux}. The atmospheric neutrino background event distribution is also shown for comparison (grey).}
    \label{fig:ngc1068_ic_events}
\end{figure}

\begin{figure}[t]
    \centering
    \includegraphics[width=\columnwidth]{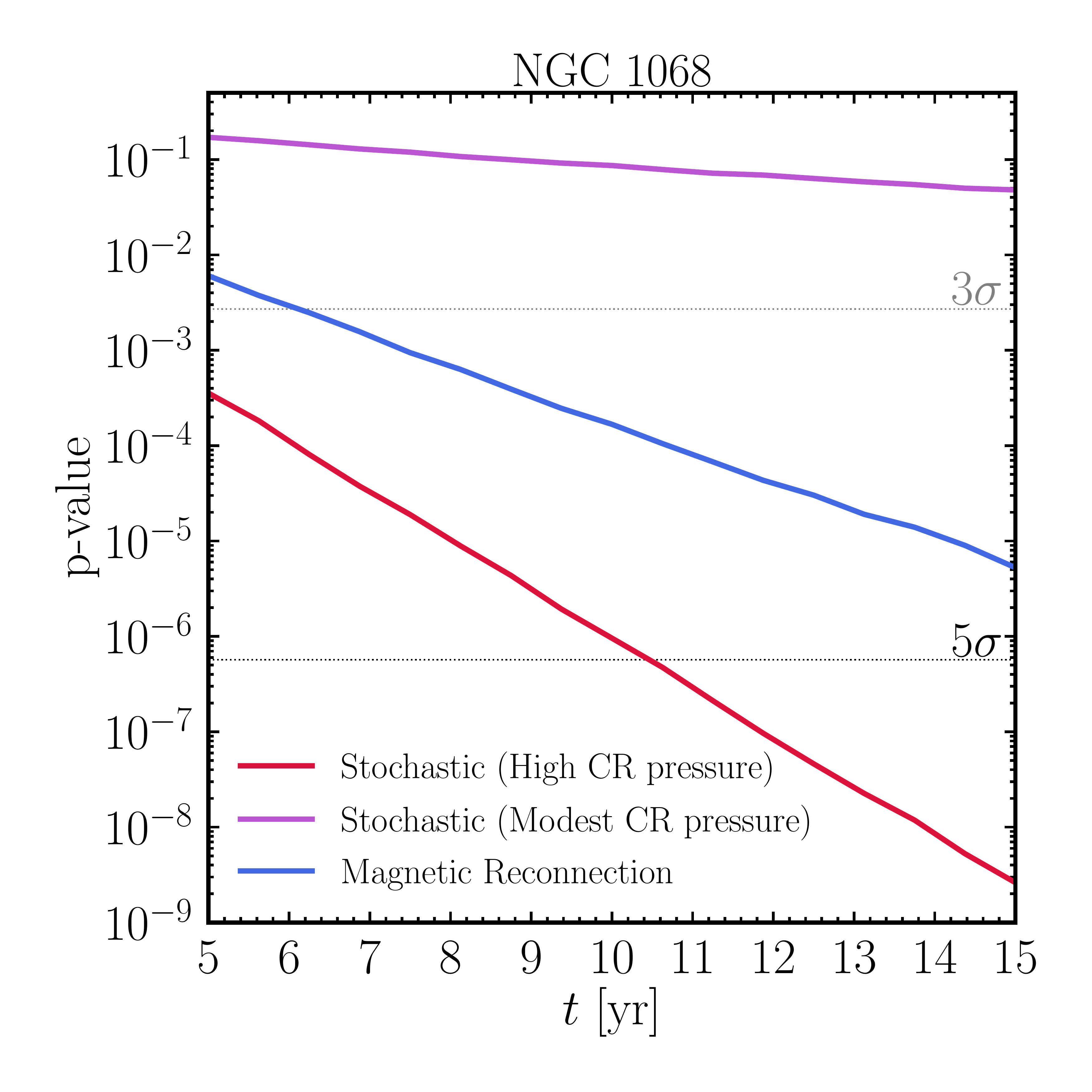}
    \caption{Prospects for identification of neutrino emission from NGC 1068. {Local p-values} are shown for stochastic with high CR pressure (red), stochastic with modest CR pressure (violet), and reconnection (blue) scenarios considered in this study.}\label{fig:ngc1068_ic_pval}
\end{figure}

Finally, we consider the {\em magnetic reconnection} scenario for particle acceleration. In this case, the neutrino flux approximately follows mainly the initial CR spectrum until the $p\gamma$ process becomes the dominant channel for the production of pions. Therefore, this scenario leads to the spectrum close to a power-law spectrum with a cutoff at high energies. For the injected CR spectrum, we assume a spectral index of 2. { The normalization and CR maximum energy are set such that the modeled flux is constrained to the IceCube steep spectrum reported for NGC 1068 while the $P_{\rm CR}/P_{\rm th}$ is bound to be smaller than 0.5. We find $E_p^{\rm rec} \approx 5 \, \rm PeV$ for this purpose. Smaller values of $E_p^{\rm rec} $ cannot accommodate the IceCube flux without violating the CR to thermal pressure maximum band. Larger values, on the other hand, would create an excess at high energies that are disfavored by the steep spectra reported for NGC 1068.  As described in Sec. \ref{sec:model}, we set $\eta_{\rm acc}=300$ for magnetic reconnection acceleration. For NGC 1068, $E_p^{\rm cool}$ is too high to match the IceCube data.}

Figure~\ref{fig:ngc1068nuflux} shows the three modeled neutrino fluxes from NGC 1068. We also projected the best-fit spectrum reported by the IceCube Collaboration. The best-fit power-law spectrum corresponds to the $\sim$ 51 excess neutrinos found from the direction of NGC 1068. The shaded area shows the uncertainty on the fitted spectrum as reported by IceCube. As shown, all modeled neutrino spectra are within the 68\% uncertainty of the measured spectrum. The parameters that we adapt in each scenario for particle acceleration and interaction efficiency are presented in Table~\ref{tab:params}. The common parameters among different scenarios are the same as \cite{Murase:2019vdl}.
The injected CR, i.e., proton, differential luminosity for the three scenarios shown in Fig.~\ref{fig:ngc1068nuflux} is presented in the Appendix (see Fig.~\ref{fig:ngc1068CRLum}).

We should note that a single power-law spectrum is not a realistic spectral energy distribution for neutrino emission from individual astrophysical objects. While neutrino and $\gamma$-ray spectra may, in general, reflect the initial CR spectrum, the shape of neutrino and $\gamma$-ray fluxes depends on the nature of the interaction, thresholds, and the opacity of the source. The neutrino spectra provided in this study take all these into account. On the other hand, the diffuse flux of high-energy neutrinos (or $\gamma$-rays) over a specific range of energies may be explained by a power-law since the superposition of the individual sources would wash out the features.

We use the modeled neutrino spectra for NGC 1068 to compare with the findings of the IceCube 10 yr point source study. In addition, we investigate the prospects for identification of each neutrino emission scenario in the next decade of IceCube operation. 

\begin{table*}[t]
\begin{center}

\begin{tabular}{l||c|cc|ccc}
\hline
\hline
Model & $P_{\rm CR}/P_{\rm th}$ ($\epsilon_{\rm CR}$) & $q$ & $\eta_{\rm tur}$  &  $s$  &  $\eta_{\rm acc}$ & $E_p^{\rm rec}$ \\
\hline
Stochastic acceleration with high CR pressure & 0.5 (0.03) & 5/3 & 50 &  - & - & -\\
Stochastic acceleration with modest CR pressure & 0.008 (0.0009) & 5/3 & 10  & - & - & -\\
Magnetic Reconnection & 0.5 (0.01) & - & - & 2.0 & $300$ & 5 PeV \\
\hline
\end{tabular}
\caption{Model-dependent parameters and quantities adapted for estimating the neutrino flux from NGC 1068 in each scenario of particle acceleration shown in Fig.~\ref{fig:ngc1068nuflux}. \label{tab:params}}
\end{center}
\end{table*}

In order to find the p-value for observation of neutrinos from NGC 1068 over the background of atmospheric neutrinos, we calculate the number of signal neutrinos using the publicly available effective area for the IceCube point source selection \citep{Aartsen:2016oji}. We also estimate the expected number of background atmospheric neutrinos using the zenith dependent atmospheric neutrino flux reported by \cite{Honda:2006qj}, assuming a resolution of 0.7$^\circ$. This is indeed larger than IceCube's nominal angular uncertainty of 0.5$^\circ$ at high energies ($\gtrsim 100$ TeV). We chose this value because most neutrinos in our predicted spectra are found at the 1-10 TeV range. The energy distribution of the events per year corresponding to the three neutrino scenarios we consider here is shown in Fig.~\ref{fig:ngc1068_ic_events}.

\begin{figure}[t]
    \centering
    \includegraphics[width=\columnwidth]{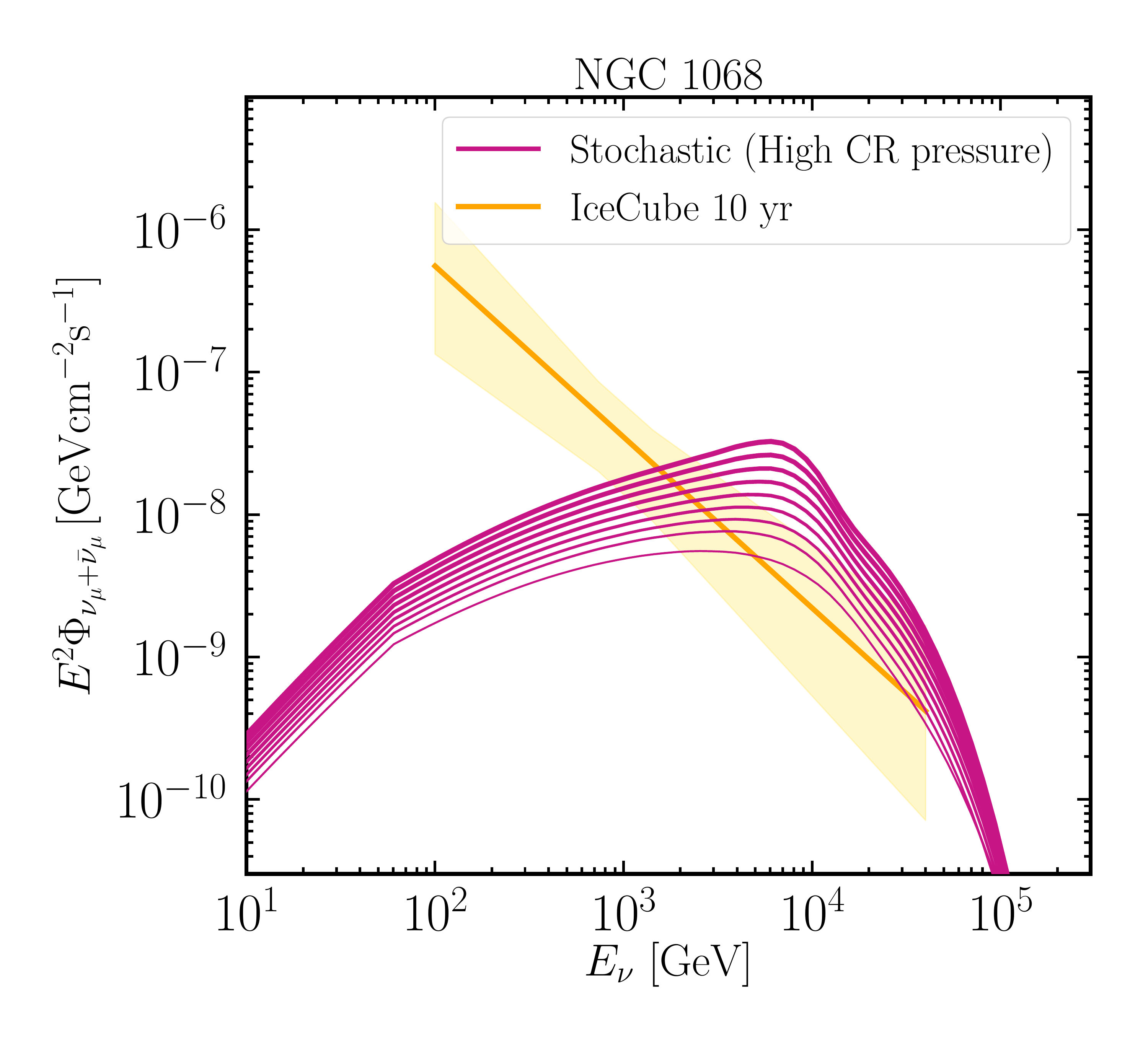}
    \caption{Neutrino spectrum for NGC 1068, assuming different values of the X-ray luminosity. Here, we study how the neutrino spectrum depends on the uncertainties associated with X-ray observations. The line shows expected neutrino emission for the stochastic acceleration scenario, for $L_{\rm X}$ between $10^{42.8}~{\rm erg}~{\rm s}^{-1}$ and $10^{43.8}~{\rm erg}~{\rm s}^{-1}$.}
    \label{fig:ngc1068-LXS}
\end{figure}

The energy distribution of the events is a convolution of the spectrum and the IceCube effective area in the direction of NGC 1068. The stochastic acceleration scenario with the high CR pressure case results in an event distribution that peaks around 10 TeV. The modest CR pressure case for the stochastic acceleration yield a lower rate, and the majority of the events are found with energies around 30 TeV. On the other hand, the harder spectrum at higher energies in the magnetic reconnection scenario leads to a relatively flat distribution of events and creates an excess beyond 100 TeV where atmospheric neutrinos are scarce. 

We present the p-value\footnote{Note that the p-values presented here are pre-trial local p-values and do not account for the trials associated with point source studies.} for the observation of each scenario in 5 to 15 years of operation of IceCube in Fig.~\ref{fig:ngc1068_ic_pval}. We use the method described in \cite{ATLAS:2011tau} to estimate the statistical significance for identification of the neutrino emission; for details, see Appendix. {The stochastic acceleration scenario with the high CR pressure, compatible with the IceCube 10 year flux measurements, provides the most likely scenario for observation of NGC 1068 in IceCube, reaching to 5$\sigma$ level in less than 12 years of data. This is in accordance with the reported local p-value in IceCube 10 year point source study.} {The magnetic reconnection scenario is expected to be identified at better than 3$\sigma$ within the 10 years, while near a decade more observation might be required to establish a significance at the discovery level. Finally, the stochastic acceleration scenario with the modest CR pressure seems difficult to identify with enough significance. However, the prospects for this scenario may be conservative given uncertainties on the X-ray luminosity. As demonstrated in Fig.~\ref{fig:ngc1068-LXS}, the neutrino flux would be higher for larger intrinsic X-ray luminosities.}

The key parameter in modeling the neutrino emission from NGC 1068 is the intrinsic X-ray luminosity. The column density ($N_{\rm H}$) for NGC 1068 is very high. Therefore, it is difficult to estimate the intrinsic X-ray luminosity, and measurements often carry large uncertainties. The dominant source of uncertainty in measuring the X-ray luminosity from NGC 1068 is the fraction of X-rays that is scattered into our line of sight \citep{Janssen_2015}. X-ray luminosities as high as $10^{44}$ erg/s have been considered for NGC 1068. {\em NuSTAR} and {\em XMM?Newton} monitoring campaigns \citep{Marinucci:2015fqo} found $L_X \simeq 7^{+7}_{-4}\, \times 10^{43}$ erg/s with $N_{\rm H} \simeq 10^{25} \rm cm^{-2}$.
Here, we used a central value of $10^{43}$ erg/s, which is a conservative choice.  We further investigate the variation of the measured intrinsic X-ray luminosity for the first scenario. Figure~\ref{fig:ngc1068-LXS} shows how the spectrum changes when the X-ray luminosity is varied between $10^{42.8}$ to $10^{43.8}$ erg/s. Here, we set the CRs to thermal pressure to the maximal value of 0.5. The larger values of X-ray luminosity will increase the flux by almost an order of magnitude, while the peak of the spectrum is slightly shifted towards higher energies.
With a larger value of $L_X$, smaller values of $\eta_{\rm tur}$ and $P_{\rm CR}/P_{\rm th}$ are demanded to match the IceCube data, which may help relaxing the extreme CR production efficiency for the high CR pressure case. 

\begin{figure}[t] 
    \centering
    \includegraphics[width=1.0\columnwidth]{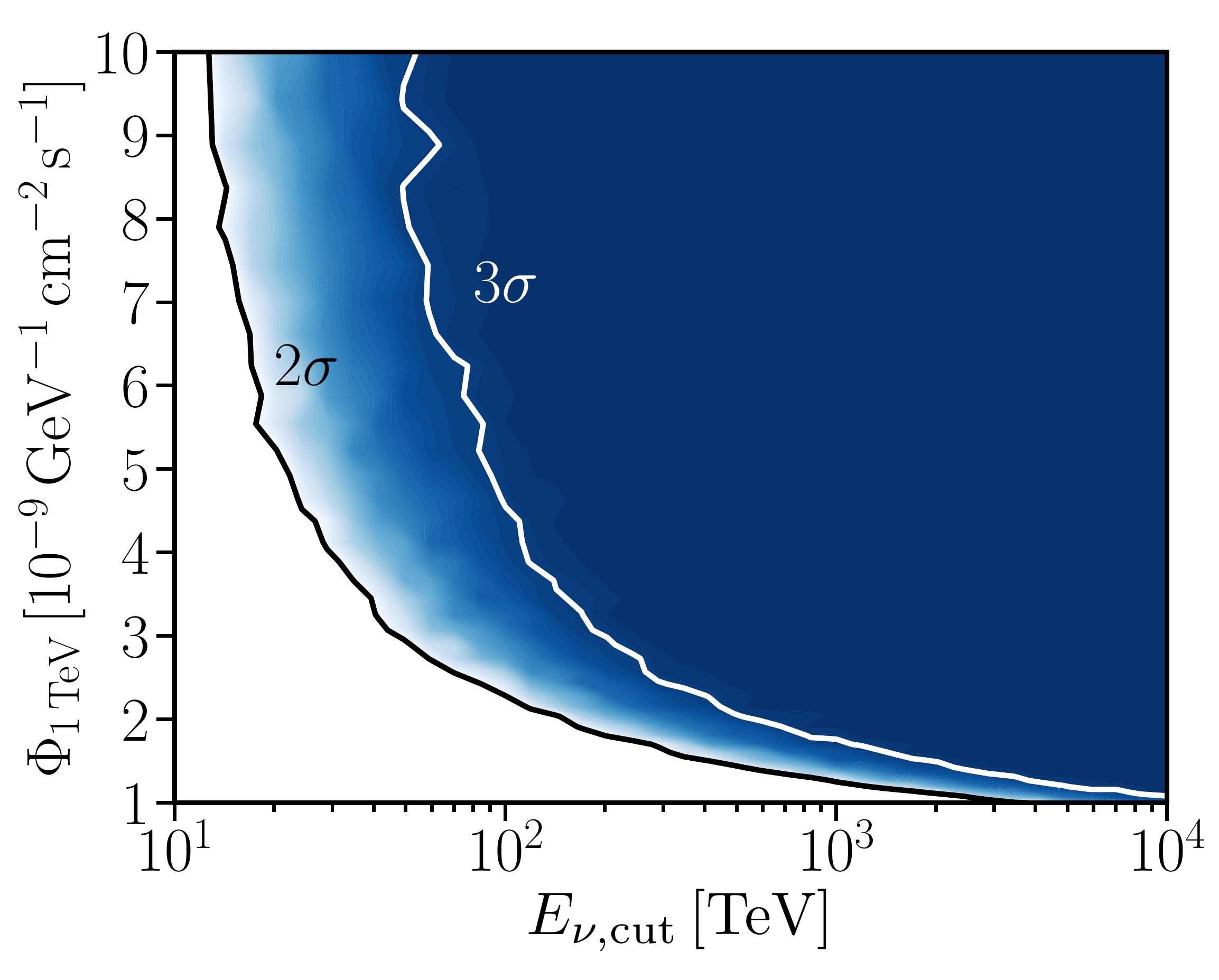}
    \caption{ Exclusion region of the parameter space of muon neutrino flux when spectrum is a single power-law ($E^{-2}$ with exponential cutoff for NGC 1068. The 2 and 3$\sigma$ confidence level exclusion regions are obtained assuming that the hard spectrum with a cutoff is not found in the direction of NGC 1068.} 
    \label{fig:cutoff}
\end{figure}

{ As we discussed earlier, in the magnetic reconnection case, the neutrino spectrum follows the CRs flat spectrum with a cutoff at very high energies. IceCube's 10 year observation disfavors a hard spectrum and suggests that the flux is suppressed at energies beyond 100 TeV. We evaluate the neutrino detection rate using a neutrino spectrum of a simple power law with an exponential cutoff to constrain the normalization and cutoff energy for the neutrino flux. We present in Fig.~\ref{fig:cutoff} the 2 and 3$\sigma$ exclusion regions for 10 years of IceCube observation. As such, magnetic reconnection scenarios with large normalization and/or large values of maximum energies are not preferred. The constraints on $E_{\nu,\rm cut}$ will also be useful for testing the accretion shock scenario. The diffusive shock acceleration theory predicts $t_{\rm acc}\approx\eta_{\rm acc}r_L/c$ with $\eta_{\rm acc}\approx(20/3){(c/V_s)}^2$, where $V_s$ is the shock velocity. This leads to $E_{p}^{\rm max}\sim {\rm a~few}~$PeV in NGC 1068~\citep{Inoue:2019yfs}. Thus, our results imply that fast acceleration models including both magnetic reconnection and accretion shock scenarios can critically be tested with near-future neutrino observations.}

\subsection{Neutrino emission from bright Seyfert galaxies}
We now use the parameters of the modeled neutrino flux from NGC 1068 to estimate neutrino spectra and prospects for observations of other bright Seyfert galaxies. We first focus on IceCube as it is currently the major operating neutrino telescope. We will extend our study into KM3NeT and the next generation of IceCube, IceCube-Gen2, in the next section. 

\begin{table*}[t]
\begin{center}
\begin{tabular}{lccccc}
\hline
\hline
Source & Declination & $z$ & $d_{\rm L}$ & Intrinsic flux & $\log$(Intrinsic luminosity)\\
 & [deg] &  & [Mpc] & [$10^{-12}\rm erg\, cm^{-2}\,s^{-1}$] & [$\rm erg\,s^{-1}$]     \\
\hline
Circinus Galaxy & -65.34 & 0.0014 &  4.2$^\star$ & 984.4 & 42.31  \\
ESO 138-1 & -59.23 & 0.0091 & 39.2 & 671.3 & 44.09  \\
NGC 7582 & -42.37 & 0.0052 & 22.4 & 507.6 & 43.48  \\
Cen A & -43.02 & 0.00136 & 3.8$^\star$ & 347.3 & 42.39 \\
NGC 1068 & -0.013 & 0.00303 & 13.0 & 268.3 & 42.93 \\
NGC 424 & -38.08 & 0.0118 & 51.0 & 188.1 & 43.77  \\
CGCG 164-019 & 27.03 & 0.0299 & 131.0 & 179.5 & 44.57 \\
UGC 11910 & 10.23 & 0.0267 & 116.7 & 157.5 & 44.41 \\
NGC 4945 & -49.47 & 0.0019 &  3.6$^\star$ & 149.4 & 41.36 \\
NGC 1275 & 41.51 & 0.0176 & 76.4 & 132.8 & 43.98 \\
 \hline
 \end{tabular}
\caption{Seyfert galaxies with the highest intrinsic X-ray fluxes in the range of 2-10 keV in BAT AGN Spectroscopic Survey (BASS) \citep{Ricci:2017dhj}. Here, we show the declination, redshift, and the intrinsic X-ray fluxes and luminosities reported by BASS in 2-10 keV band. For sources closer than 10 Mpc, specified by $\star$, we employ distances reported by astronomical measurements. For other sources, the luminosity distances are evaluated from the reported redshift.}
 \label{tab:sources}
\end{center}
\end{table*}

For this purpose, we rank Seyfert galaxies by their intrinsic X-ray fluxes measured by BAT AGN Spectroscopic Survey \citep{Ricci:2017dhj}. We select the 10 brightest sources for our study and extract their intrinsic X-ray luminosities from the BASS catalog. Table \ref{tab:sources} presents the list of sources and their declination, distance, redshift, intrinsic X-ray flux, and luminosity in 2-10 keV band. 
In order to estimate the luminosity distance for source, we incorporate the redshift reported in BASS together with the cosmological parameters introduced in the Introduction, with the exception of Cen A, Circinus Galaxy, and NGC 4945, for which we  employed astronomical measurements for evaluation of the distance as reported by \cite{Harris:2009wj, 2013AJ....145..101K,2013AJ....146...86T}. We should note that the neutrino flux is insensitive to the distance, because the neutrino flux is roughly proportional to the X-ray flux.

\begin{figure*}[t]
    \centering
    \includegraphics[width=\textwidth]{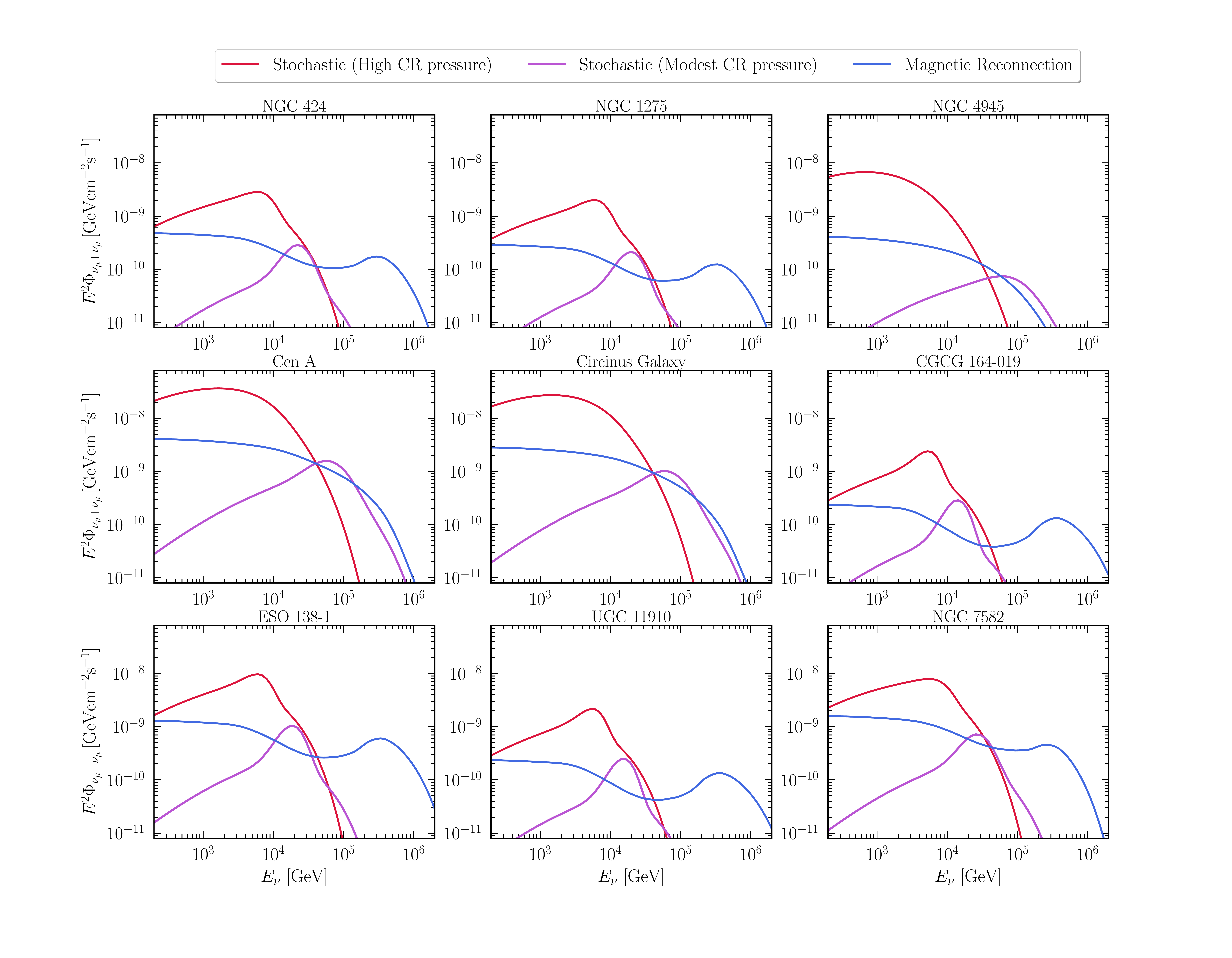}
    \caption{Neutrino flux for bright Seyfert galaxies considered in this study. Here, we show the high (red) and modest (purple) CR pressure stochastic acceleration scenarios as well as the magnetic recoonection scenario (blue), which provide compatible fluxes with the best-fit flux for NGC 1068 or the total neutrino spectrum measurement for parameters presented in Table~\ref{tab:params}.
    }\label{fig:nuflux-all}
\end{figure*}

We implement the parameters that we adopted in the previous section for modeling the neutrino spectrum from NGC 1068, listed in Table~\ref{tab:params}. Again, we consider 3 scenarios: the stochastic acceleration scenario compatible with the 10-year IceCube observation of NGC 1068 (high CR pressure), the stochastic acceleration scenario compatible with the medium-energy excess in the diffuse neutrino flux (modest CR pressure), and the magnetic reconnection scenario. Similar to NGC 1068, we impose the physical constraint on the ratio of CR pressure to the thermal pressure to 0.5. Therefore, the normalization of the CR spectrum is adjusted to maintain this bound whenever this ratio exceeds 0.5 with NGC 1068 parameters. Figure~\ref{fig:nuflux-all} demonstrates the spectrum for the source list considered in this study. For the stochastic acceleration scenario with the high CR pressure, we have adjusted flux to match the ratio of CR pressure to the thermal pressure to 0.5 by rescaling $f_{\rm inj}$, which are used in our analyses hereafter. The measured intrinsic X-ray luminosity and the distance of the source define the shape and magnitude of the spectra. In the stochastic acceleration scenario, sources with $L_X \gtrsim 10^{43.5}\rm~erg~s^{-1}$ peak at relatively higher energies with a narrow width while sources with smaller intrinsic X-ray luminosity demonstrate a broader spectrum peaking at lower energies. In this scenario, Cen A has the highest level of flux in the list of bright Seyfert galaxies. For the stochastic acceleration with modest CR pressure, we employ $f_{\rm inj}$ value that yields the CR to thermal pressure of $\sim$ 0.008 for NGC 1068. That corresponds to $f_{\rm inj}\simeq 10^{-7} (5\times10^{-5})$ at $2m_pc^2$ ($1000m_pc^2$). Finally, in the magnetic reconnection scenario, the larger X-ray luminosity leads to a larger contribution of photohadronic processes. { For each source, as the maximum energy in this scenario depends on the size of the system as well as the cooling effects. We evaluate $E_p^{\rm rec} $ by rescaling with $M_{\rm BH}$ as $E_p^{\rm rec}\approx E_p^{\rm rec-NGC \,1068} (M/M_{\rm NGC\, 1068})^{1/2}$. This scaling is motivated by the model assumptions with $l_{\rm rec}/(2GMc^{-2})$ and $\beta$ constant.}

The majority of bright nearby Seyfert galaxies in Table \ref{tab:sources} are located in the Southern hemisphere, as pointed out by \cite{Murase:2019vdl}. IceCube's event selection is optimal for the Northern sky, where the Earth acts as a shield for the atmospheric muons. In the Southern hemisphere, the event selection imposes a higher-energy threshold on the energy of the neutrinos to suppress the atmospheric muon background.  This feature suppresses the event rate for the majority of the luminous Seyfert galaxies. We show the expected events from the sources in this list in Fig.~\ref{fig:event_ic_all} in Appendix. Except for the sources NGC 1275, UGC 11910, and CGCG 164-019 that are in the Northern hemisphere, the event rates for the rest of the sources are low, weakening the likelihood of identifying individual sources in IceCube.

Using the expected signal and background rates, we estimate the likelihood for observations of these sources in IceCube. Table~\ref{tab:icpvals} summarizes the expected p-values under each emission scenario for 10 years of IceCube operation. The listed p-values show that for all three acceleration scenarios, NGC 1068 is the brightest source in IceCube. While the prospects for the identification of most sources are not much promising due to the suppression of events in the Southern hemisphere, with continued collection data CGCG 164-019 and NGC 1275 are likely to be observed at 3$\sigma$ level in 20 years of IceCube operation. However, we should note that the likelihood of observations depends on the neutrino emission scenario: the stochastic acceleration scenario with the high CR pressure, compatible with NGC 1068 parameters, would yield $\sim 3\sigma$.

Another source in the list worth discussing is NGC 4945. The IceCube 10 year analysis found an excess of $\sim$ 1 event in its direction, corresponding to a p-value of 0.48, which is consistent to the expectations found in our study. NGC 4945 is a starburst galaxy. We should note that similar to NGC 1068, the neutrino flux from this source cannot be explained by the starburst scenarios such as \citet{Eichmann:2015ama}.

Other than NGC 1068, our predictions indicate that even optimistic scenarios are not strong enough to yield a statistically significant measurement of the neutrino emission from the rest of the bright nearby sources. As such, a stacking search for neutrino emission from the bright Seyfert galaxies is going to offer the best chance for identifying these sources in IceCube. Stacking analyses are widely used to study the correlation of the arrival direction of high-energy neutrinos and a catalog of sources. The cumulative signal from all the sources improves the sensitivity and makes it possible to examine a class of sources with lower fluxes.

\begin{table*}[t]
\begin{center}
\begin{tabular}{lccr}
\hline
\hline
  & &  p-value & \\
 Source & Stochastic  (High CR pressure) & Stochastic (Modest CR pressure) & Magnetic reconnection\\
\hline
NGC 1068 & $10^{-6}$ & 0.09 & 1.8$\times10^{-4}$\\
NGC 1275 & 0.03 & 0.3 & {0.1}  \\
CGCG 164-019 & 0.04 & 0.3 & { 0.1} \\
UGC 11910 & 0.1 & 0.4 & { 0.09} \\
Cen A & 0.5 & 0.2 & { 0.2} \\
Circinus Galaxy & 0.5 & 0.3 & { 0.3} \\
NGC 7582 & 0.5 & 0.5 & { 0.1} \\ 
ESO 138-1 & 0.5 & 0.5 & { 0.09} \\
NGC 424 & 0.5 & 0.5 & 0.5 \\
NGC 4945 & 0.5 & 0.5 & 0.5\\
\hline
 \end{tabular}
\caption{Prospects for observations of bright nearby Seyfert galaxies in 10 years of IceCube operations.} 
 \label{tab:icpvals}
\end{center}
\end{table*}

\begin{figure}[t] 
    \centering
    \includegraphics[width=1.0\columnwidth]{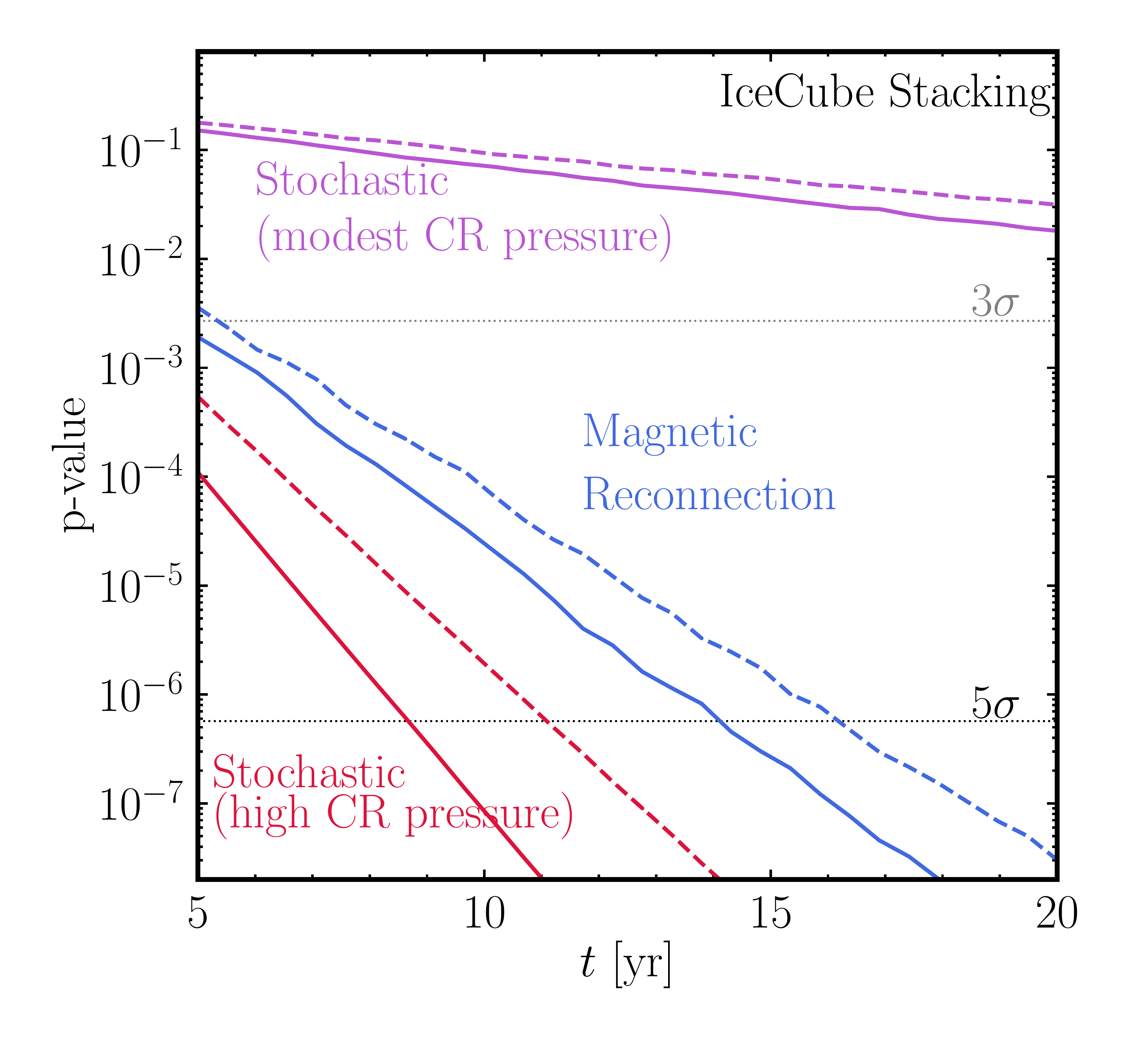}
    \caption{Projected p-values for stacking of 10 brightest Seyfert galaxies in 5 to 20 years of IceCube operation, for the three acceleration scenario considered in this study. The solid lines show the p-values for the 10 bright sources listed in Table~\ref{tab:sources}. The dashed line presents the prospects when the emission from Cen A and NGC 1275 is excluded.}
    \label{fig:icecube_stack}
\end{figure}

We project the prospects for identifying each scenario of neutrino emission in a stacking search for 5 to 20 years of IceCube in Fig.~\ref{fig:icecube_stack}. The solid lines show the p-value for the duration of IceCube observation for each scenario. The most promising scenario is the stochastic acceleration scenario with the high CR pressure case. The high-level neutrino flux in this scenario would result in an excess of signal that can reach a discovery level of 5$\sigma$ in less than 10 years of IceCube operation. The hard spectrum in the magnetic reconnection scenario would result in an excess distinguishable from the background at high energies that can yield a 3$\sigma$ evidence in about 7 years of IceCube. The significance can reach a discovery level of 5$\sigma$ within the 20 years of operation for this scenario.

The most conservative scenario here is the stochastic acceleration scenario with the modest CR pressure. This scenario would not yield a signal at a significant level with the current generation of IceCube. Mainly due to the fact that most sources are in the Southern hemisphere. We investigate the likelihood of identifying this scenario in the next generation of neutrino telescopes in the next section.

{ We should note that the significance for the stacking analysis for stochastic acceleration with high CR pressure scenario is dominated by the neutrino flux in the direction of NGC 1068. However, it is worth mentioning that for this scenario, a stacking analysis of nearby bright Seyfert galaxies when NGC 1068 is excluded is expected to identify neutrino emission at the level of 3$\sigma$ with 15 years of IceCube data. Such analysis can test this scenario independently, without relying on the neutrino emission reported from the direction NGC 1068, which this scenario is based upon.}

Our predicted neutrino emission was built upon the disk-corona model of AGN. Two of the sources in the list of bright Seyfert galaxies, Cen A and NGC 1275, are seen with a high jet activity and it is likely that the X-ray emission arises from the jet rather than the disk. We will further investigate the prospects for observations of bright Seyfert galaxies if the disk-corona emission would not be dominant. The dashed lines in Fig.~\ref{fig:icecube_stack} show the p-value when Cen A and NGC 1275 are not considered in the source list. While the likelihood of observation is decreased, the cumulative neutrino emission is strong and the stacking analysis of the rest of the bright Seyfert galaxies can still reach a significant level in the lifetime of the IceCube detector.

\begin{figure}[t]
    \centering
    \includegraphics[width=\columnwidth]{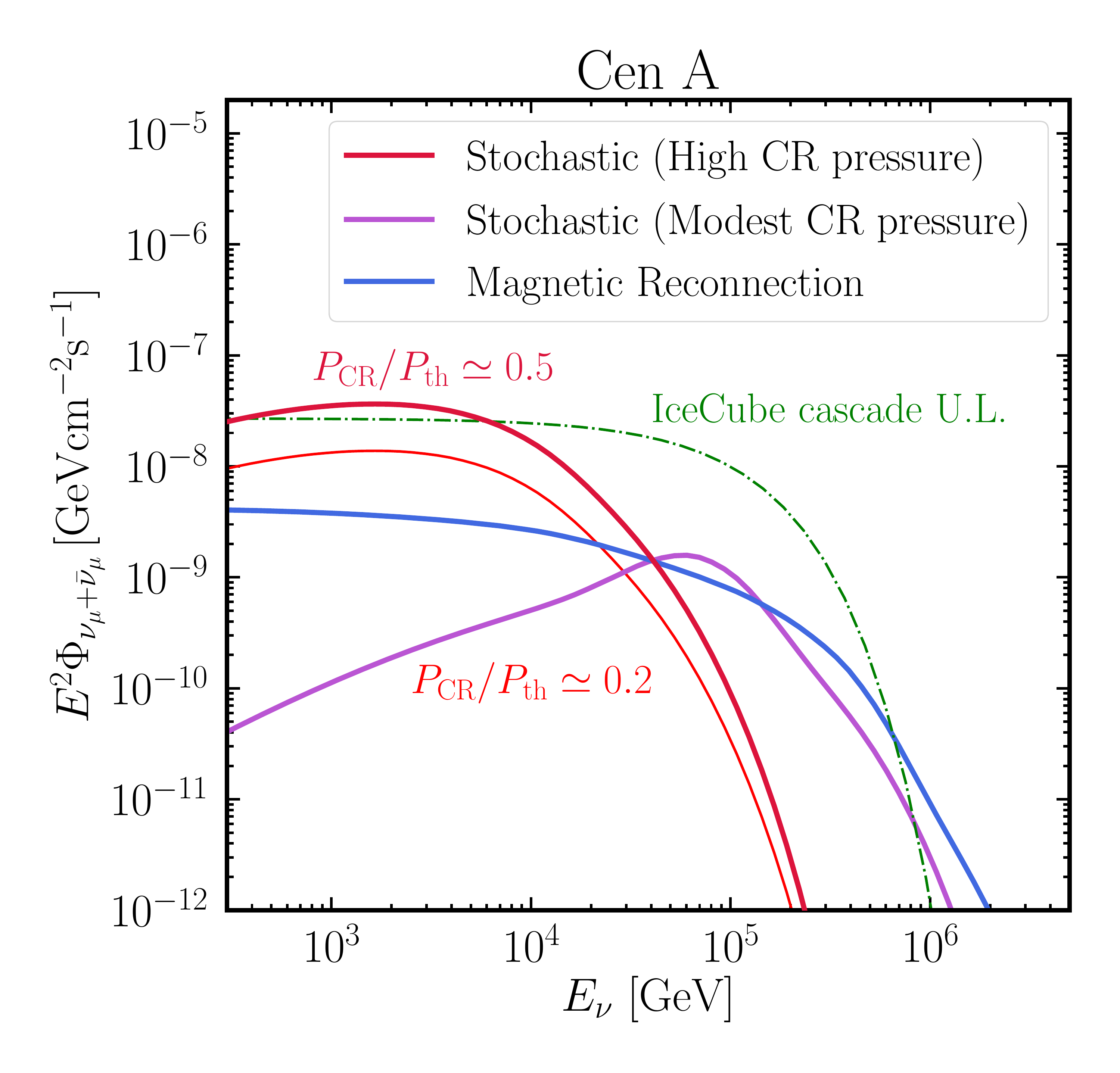}
    \caption{Neutrino spectra from Cen A compared with the discovery potential of a joint IceCube cascade source search for the sources in the Southern sky \citep{Aartsen:2020xpf}.}
    \label{fig:cenAul}
\end{figure}

For Cen A, we should note that although one zone models typically attribute the high-energy to the jet~\cite{Falcone:2010fk}, it may be difficult to explain some of the X-ray properties with such scenario~\citep{Tachibana:2015ita}. The observed soft lags in X-rays from Cen A may indicate their coronal origin. We should note that despite the large magnitude of the flux predicted under both stochastic acceleration and magnetic reconnection scenarios for Cen A in our study, our prediction is compatible with current upper limits imposed by the ANTARES and IceCube Collaboration. In Fig.~\ref{fig:cenAul} we compare the predicted neutrino flux with the current limits from the IceCube cascade source search in the Southern sky \citep{Aartsen:2020xpf}. The upper limit is obtained for a power-law with exponential cutoff, which is the closest shape to the predicted spectra in our study. 

In summary, observations of stacked neutrino emission from nearby, bright Seyfert galaxies is promising with IceCube, which could reveal the dominant sources responsible for the medium-energy excess in the spectrum of high-energy cosmic neutrinos. The likelihood for observations of these sources will be enhanced by the improvements in the event selection in the Southern hemisphere. We will address this in Sec.~\ref{sec:summary}.

\section{Future neutrino telescopes}\label{nextgen}
In this section, we explore the prospects for identifying bright nearby Seyfert galaxies in the next generation of neutrino telescopes. The major development in the high-energy neutrino astrophysics would be driven by KM3NeT, currently under construction in the Mediterranean, and the next generation of IceCube: IceCube-Gen2.

KM3NeT is a cubic km-scale water Cherenkov detector, which is designed to enhance the exposure in the Southern hemisphere. The larger effective area and improved angular resolution, compared to the current neutrino detector in the Mediterranean, ANTARES, are expected to provide better sensitivity to the sources in the Southern sky~\citep{Adrian-Martinez:2016fdl}. This advancement would enhance the potential of identifying neutrino emission from bright Seyfert galaxies as the majority of them are located in the Southern hemisphere. 

\begin{figure}[t]
    \centering
    \includegraphics[width=\columnwidth]{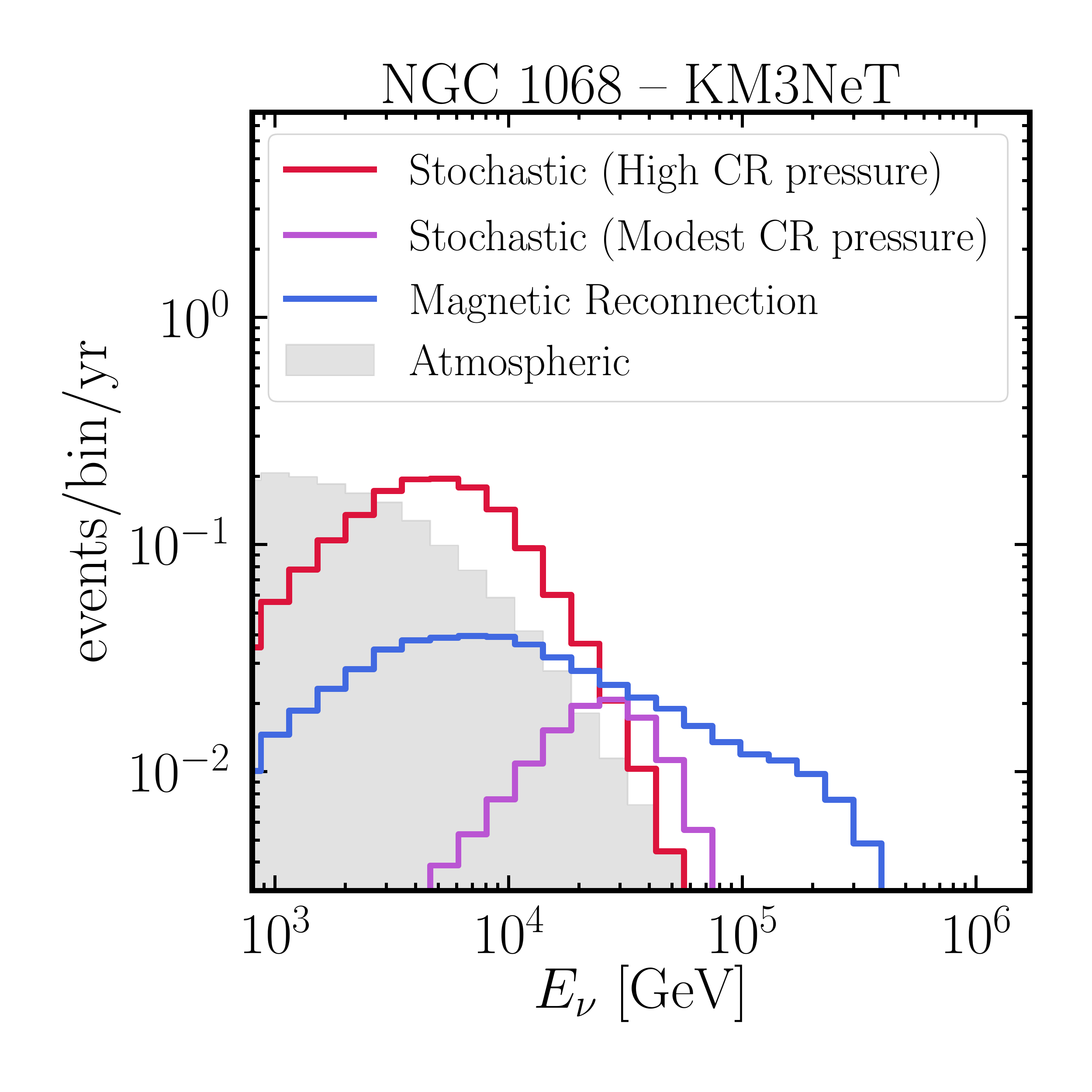}
    \caption{Expected signal and background events in KM3NeT neutrino spectra presented in Fig.~\ref{fig:ngc1068nuflux}. }
    \label{fig:ngc1068_events_km3net}
\end{figure}

In order to estimate the likelihood of identifying neutrino emission, we incorporate the effective area for upgoing muon neutrinos in KM3NeT. We use the effective area reported in \citep{Adrian-Martinez:2016fdl} and include the trigger efficiency for background rejection in point source analyses. The publicly available effective area for the two blocks of ARCA is averaged over the zenith covering the upgoing events. Therefore, in our estimation of the atmospheric background events, we use the average atmospheric neutrino flux instead of zenith dependent ones that we used for IceCube.  
We estimate the signal and background events from the sources listed in Table~\ref{tab:sources}. Unlike IceCube, KM3NeT is not located at the geographic pole and sources' zenith angle vary. Thus, we need to take into account the duration that the source is positioned below the horizon. As such, we take into account the visibility of the source when calculating the event rate and evaluating the prospects for their identification. 

We first examine the likelihood for observation of NGC 1068 in KM3NeT. Figure~\ref{fig:ngc1068_events_km3net} shows the expected signal and background rate.
Given that the source visibility is 50\%, it would not be identified in KM3NeT, even for the most optimistic scenario, i.e, the stochastic acceleration with the high CR pressure which would yield a p-value of few percent after 5 years.

\begin{figure}[t]
    \centering
    \includegraphics[width=\columnwidth]{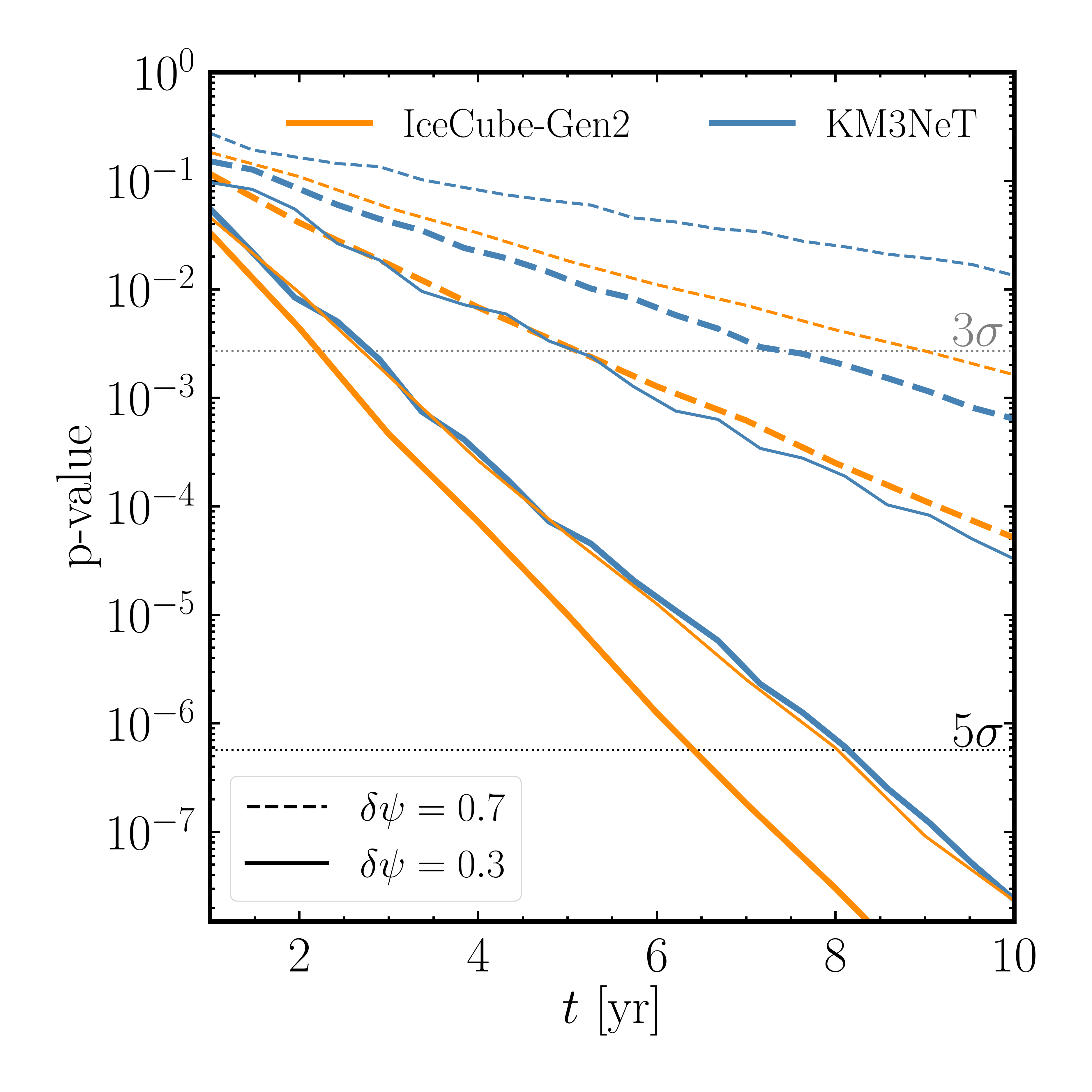}
    \caption{Prospects for observation of the bright Seyfert galaxies in the next-generation neutrino telescopes: KM3NeT and IceCube-Gen2. The solid (dashed) lines show expectations for 0.3$^\circ$ (0.7$^\circ$) angular resolution for the modest CR pressure scenario. The tick lines show the prospects for identification of the 10 nearby bright sources in Table~\ref{tab:sources} in a stacking analysis. The thin lines show the prospects for identification of the sources in the absence of a signal from disk-corona model from Cen A and NGC 1275.}
    \label{fig:pval_gen2_km3net}
\end{figure}

\begin{table*}[t]
\begin{center}
\begin{tabular}{lcccr}
\hline
\hline
  & & &  p-value 1 yr (3 yr)& \\
 Source & Visibility & Stochastic (high CR pressure) & Stochastic (Modest CR pressure) & Magnetic Reconnection\\
\hline
Cen A & 0.7 & 0.001 (9.3$\times 10^{-8}$) & 0.2 (0.07) & {0.2 (0.05)}\\
Circinus Galaxy & 1.0 & 0.008 (1.9$\times 10^{-5}$) & 0.2 (0.09) & {0.2 (0.07)}\\
ESO 138-1 & 1 & 0.1 (0.02) & 0.4 (0.3) & { 0.3 (0.08)}\\
NGC 7582 & 0.7 & 0.2 (0.04) & 0.4 (0.3) & 0.4 (0.2) \\
NGC 1068 & 0.5 & 0.2 (0.05) & 0.4 (0.4) & 0.4 (0.2) \\
NGC 4945 & 0.8 & 0.5 (0.2) & 0.5 (0.4) & 0.5 (0.4) \\
NGC 424 & 0.7 & 0.4 (0.2) & 0.5 (0.4) & 0.5 (0.4) \\
UGC 11910 & 0.5 & 0.4 (0.4) & 0.5 (0.5) & 0.5 (0.5) \\
CGCG 164-019 & 0.4 & 0.4 (0.3) & 0.5 (0.5) & 0.5 (0.5) \\
NGC 1275 & 0.3 & 0.4 (0.4) & 0.5 (0.5)  & 0.5 (0.5)\\
 \hline
 \end{tabular}
\caption{Prospects for observation of nearby bright Seyfert galaxies in one years of KM3NeT observations.}
 \label{tab:km3netpvals}
\end{center}
\end{table*}

We present the p-values for observations of all sources in 1 and 3 years of operation of KM3NeT in Table~\ref{tab:km3netpvals}. The brightest source in KM3NeT in the list of bright Seyfert galaxies is found to be Cen A given that its X-rays come from the coronal region (that may not be the case). Thanks to the higher level of neutrino flux, the stochastic acceleration scenario with the high CR pressure is going to be found at 3$\sigma$ in the first year of observation, and the significance can reach 5$\sigma$ with additional two years of observation. 

Circinus Galaxy is found to be the second brightest source in the Southern sky. In addition to its high level of neutrino flux at TeV energies, the source benefits from a 100\% visibility in KM3NeT. Therefore, the likelihood for its observation is high, which can exceed 3$\sigma$ in 3 years of operation for the stochastic acceleration scenario with the high CR pressure.

As the signal events from the rest of the sources in the list fall short of yielding a statistical significance in 3 years, we now turn into the prospects for observation of neutrino emission in a stacking analysis. We only consider the modest CR pressure scenario in stochastic acceleration since emission under either of the other two scenarios should be identified by IceCube. In addition to KM3NeT, we consider IceCube-Gen2 for the stacking search in this scenario. Here, we assume that the effective area for IceCube-Gen2 is $\sim$ 5 times larger than the current IceCube detector.

We present the p-values expected for modest CR pressure stochastic acceleration scenario for KM3NeT together with the ones for IceCube-Gen2 in Fig.~\ref{fig:pval_gen2_km3net}. We project the prospects for identification of neutrino emission from the bright sources assuming an angular resolution of 0.3 (solid) and 0.7 degrees (dashed) for each detector. 
 We should note that our estimation of the prospects for identifying Seyfert galaxies are quite conservative, given that an angular resolution of 0.3 degrees or better is not that far-fetched for KM3NeT. The expected improvements in the angular reconstruction in IceCube-Gen2 will also make it easier to identify these sources.
In fact, our estimates indicate that achieving finer angular resolutions at $\sim10-30$ TeV is crucial for the identification of neutrino emission from these sources especially in the modest CR pressure case. We further show the growth of significance for a given resolution in Sec.~\ref{sec:summary}. 

\section{Discussion}

\begin{figure*}[t]
    \centering
    \includegraphics[width=\textwidth]{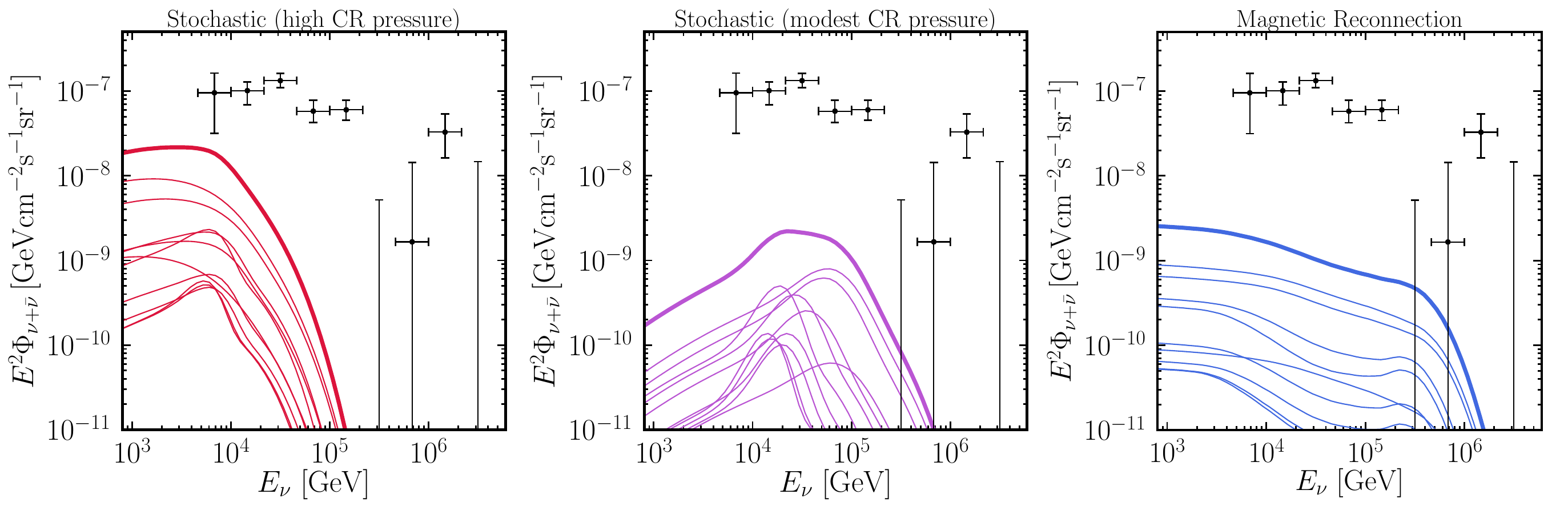}
    \caption{Accumulated {\bf all flavor} neutrino emission (thick line) and the individual neutrino flux (thin lines) for the bright nearby Seyfert galaxies considered in this study. We show the neutrino flux for stochastic acceleration with highest CR pressure (left), stochastic acceleration with modest CR pressure (middle), and magnetic reconnection (right). The data points show the measured cosmic neutrino flux in 6 years of IceCube cascade data \citep{Aartsen:2020aqd}}
    \label{fig:totalflux}
\end{figure*}

\subsection{Aggregated Fluxes}
Highly magnetized and turbulent coronae can be possible sites of particle acceleration. The system is calorimetric in the sense that sufficiently high-energy CRs are depleted via hadronuclear and photohadronic interactions.   
The high magnitude of the neutrino flux at 10--100 TeV makes this scenario a primary candidate for the medium-energy neutrino flux observed in IceCube at the level of $E_\nu^2\Phi_\nu\sim{10}^{-7}~{\rm GeV}~{\rm cm}^{-2}~{\rm s}^{-1}~{\rm sr}^{-1}$ \citep{Murase:2019vdl}. 
The diffuse flux mainly originates from AGN at high redshifts (with $z\sim1-2$), which are too far to detect as individual sources. The contribution from local sources is small, but it is still of interest to ask their aggregated flux.   

Figure~\ref{fig:totalflux} shows the individual (thin lines) and sum (thick line) of the neutrino fluxes from nearby, bright Seyfert galaxies for different acceleration scenarios considered in this study. We have divided the fluxes by $4\pi$ in order to compare with the total neutrino flux from the 6-yr cascade analysis of IceCube \citep{Aartsen:2020aqd}. Overall, each scenario predicts the contribution of the catalogued nearby sources to the total neutrino flux at 10 TeV to be within 2-10\%. 

The stochastic acceleration scenario with the modest CR pressure would mainly contribute to the 10-100 TeV region. However, the high CR pressure case would generate a significant excess of the flux below 10 TeV. This region is hard to investigate with the overwhelming flux of atmospheric neutrinos, and detailed veto techniques are required to distinguish the flux at TeV energies with a good accuracy. The magnetic reconnection scenario has the highest contribution to the flux at $\gtrsim 100$ TeV. Distinguishing this scenario from the one responsible for the flux above 100 TeV would be difficult because of the scarcity of the data at high energies. While the extension of the flux to higher energies could make it easier to identify this flux from the background, such a possibility is less motivated by the compelling signal from NGC 1068. We discuss this further later in the next subsection.

\subsection{Magnetic reconnection scenario and the diffuse neutrino flux}
\label{sec:impl}
It has been already shown that Seyfert galaxies with a CR pressure at the level 1-10\% of the thermal pressure with the virial temperature can explain the magnitude of the diffuse neutrino flux at medium energies \citep{Murase:2019vdl}. 

One of the novel points in this work is that we suggest the magnetic reconnection scenario as an alternative possibility of CR acceleration. In this scenario, we expect a power-law spectrum for CRs. The diffuse neutrino flux in this scenario is estimated to be
\begin{eqnarray}
E_\nu^2\Phi_\nu&\sim&{10}^{-7}~\frac{\rm GeV}{{\rm cm}^{2}~{\rm s}~{\rm sr}}~\left(\frac{2K}{1+K}\right){\left(\frac{10}{\mathcal{R}_p}\right)}\left(\frac{15f_{\rm mes}}{1+f_{\rm BH}+f_{\rm mes}}\right)\nonumber\\
&\times&\left(\frac{\xi_z}{3}\right)\xi_{\rm CR}{\left(\frac{L_{X}\rho_X}{2\times{10}^{47}~{\rm erg}~{\rm Mpc}^{-3}~{\rm yr}^{-1}}\right)}, \,\,\,\,\,\,\,\,
\label{eq:diffuse}
\end{eqnarray}
where $K=1(2)$ for $p\gamma(pp)$ interactions, $\xi_z\sim3$ summarizes the redshift evolution of the AGN luminosity density~\citep{Waxman:1998yy,Murase:2016gly}, ${\mathcal R}_p$ is the conversion factor from bolometric to differential luminosities, $\xi_{\rm CR}=L_{\rm CR}/L_X=(\epsilon_{\rm CR}/\epsilon_{\rm rad})(L_{\rm disk}/L_X$) is the CR loading factor defined against the X-ray luminosity, $L_{X}$ is the X-ray luminosity, $\rho_X$ is the local density of X-ray selected AGN. Therefore, the total flux is found at ${10}^{-7}~{\rm GeV}~{\rm cm}^{-2}~{\rm s}^{-1}~{\rm sr}^{-1}$.

In the magnetic reconnection scenario, we assumed an injected spectral index of 2 for the injected CRs. This assumption yields a relatively flat spectrum of high-energy cosmic neutrinos with a cutoff around 100 TeV for $\eta_{\rm acc}$ that could describe the flux reported for NGC 1068. In addition, depending on the intrinsic X-ray luminosity of the source, a significant contribution from $p\gamma$ interactions would emerge before the cutoff. 
The normalization in this scenario is set by $P_{\rm CR}/P_{\rm th} \leq 0.5$ in our analysis for NGC 1068. 
While this scenario yields a smaller flux in the $1-10$~TeV energy range, the flux becomes dominant over the background at higher energies, providing an excess that could be distinguished over the background in the sufficient years of detector operation. A higher cutoff energy in the spectrum would be therefore constrained by the lack of observation for such a hard neutrino flux from these sources. Therefore, both magnetic reconnection and shock acceleration scenarios \citep{Inoue:2019yfs} would be disfavored unless $\eta_{\rm acc}$ is fine-tuned. 

\begin{figure}[t]
    \centering
    \includegraphics[width=\columnwidth]{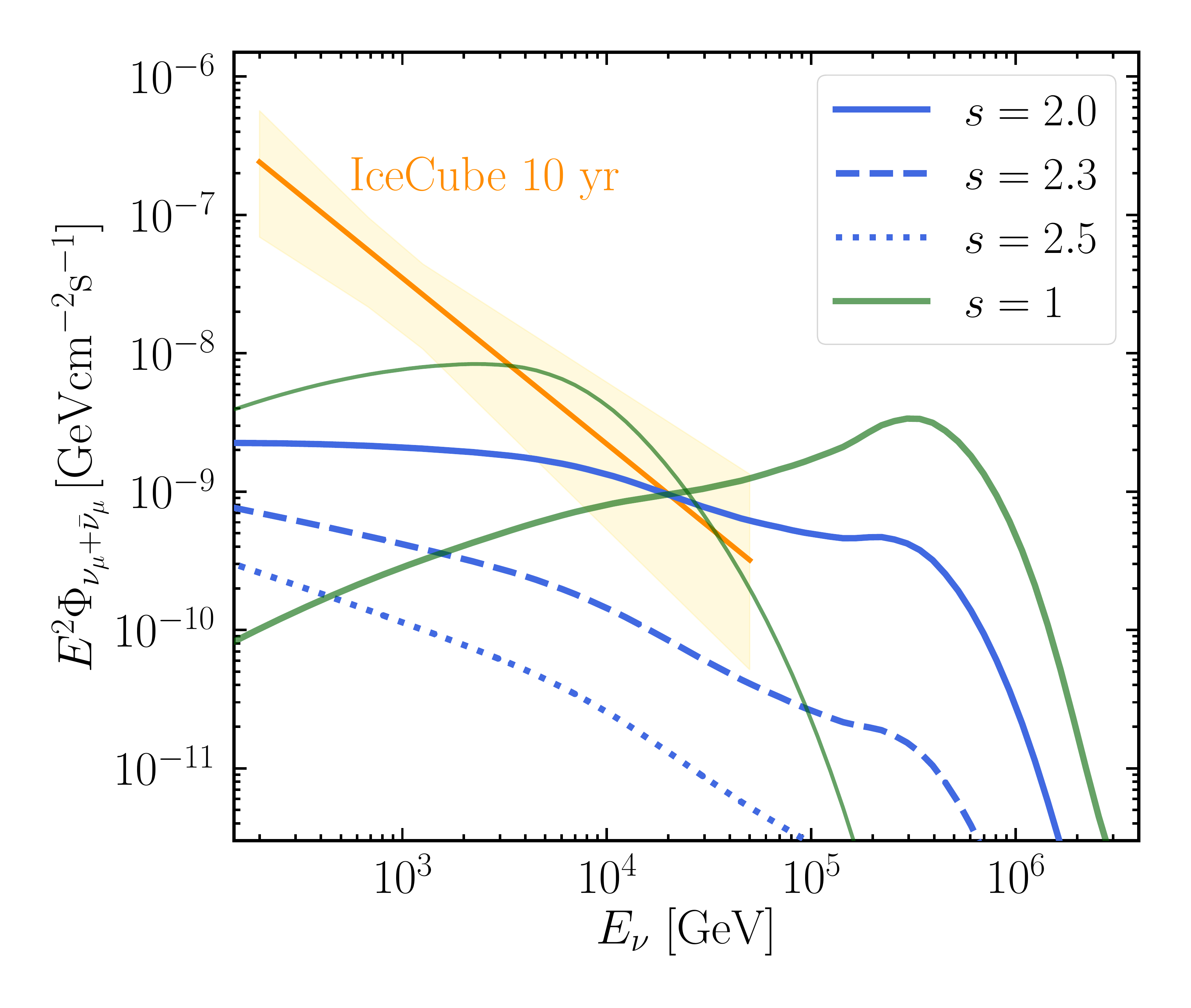}
    \caption{ Expected neutrino fluxes for different injected CR power-law in the magnetic reconnection scenario compared to the best-fit flux reported by the IceCube for NGC 1068 (orange). Here we show in blue the flux for spectral indices -2, -2.3, and -2.5.  For all these scenarios $\eta_{acc}=3 \times 10^2$ and $E^{\rm rec}_p=5\, \rm PeV$. In addition, we show the flux for spectral index of -1, with $E^{\rm rec}_p=5\, \rm PeV$ (thick green) and $E^{\rm rec}_p=0.1\, \rm PeV$ (thin green). For both cases, $\eta_{acc}=3 \times 10^2$.}
    \label{fig:ngc1068-indices}
\end{figure}

We further investigate the expected flux in this scenario by considering a steeper CR spectrum. In Fig.~\ref{fig:ngc1068-indices}, we show the neutrino flux corresponding to injected CR indices of 2.3, and 2.5 in addition to spectral index of 2 that we used before. As can be seen, when restricting the pressure of CRs to less than 50\% of the thermal pressure, it is going to be difficult to explain the measured flux NGC 1068 by the IceCube Collaboration. { We should note that although a spectral index of $\sim 2.5$ or softer is motivated for CR acceleration in magnetic reconnection \citep{Ball_2018}, such index is obtained for electrons, not protons, and it has not been supported by simulation thus far. For magnetic reconnections, harder spectra with index $\sim 1$ may be motivated, see e.g., \citep{PhysRevLett.114.061101, Guo_2016}. However, this case would induce a high $\sigma$ regime and a super-relativistic plasma, which is not compatible with our modeling that considers the low $\sigma$ regime. {Nevertheless, we have shown the case for $s=1$ in Fig.~\ref{fig:ngc1068-indices} for parameters that we adapted for NGC 1068 in Sec. \ref{sec:NGC1068}. As shown, in this scenario, we can explain the IceCube data if we change the maximum energy.}}

\section{Summary \& Outlook}\label{sec:summary}
In this study, we presented promising neutrino emission scenarios from nearby, bright Seyfert galaxies and quasars, which are typically classified as radio-quiet AGN. Our predictions were built upon the disk-corona model of AGN. \cite{Murase:2019vdl} already demonstrated that this scenario can explain the medium-energy flux of high-energy cosmic neutrinos observed in IceCube, and provide a consistent multi-messenger picture because such sources are opaque to very high-energy $\gamma$-ray emission.
Here, we presented the neutrino emission from the individual bright Seyfert galaxies by incorporating their measured intrinsic X-ray luminosities. We considered three scenarios for particle acceleration and evaluated the prospects for identifying the neutrino emission in both point source and stacking searches.

We showed that NGC 1068 is the brightest source in IceCube and can explain the near 3$\sigma$ observation reported by the 10 year analysis of IceCube data \citep{Aartsen:2019fau}. 
We found that, with a slightly higher value for the pressure ratio compared to what was previously considered \citep{Murase:2019vdl}, the model could describe the steep spectra observed for neutrino emission from NGC 1068 as well as the reported significance for the number of signal events. {Our projection demonstrates that the identification of high-energy neutrino emission from NGC 1068 at the level of 5$\sigma$ is likely  when the optimistic scenario of the maximal CR pressure to the thermal pressure is assumed.} For less optimistic scenarios, where a moderate value of the pressure ratio is considered, the significance at the level of 3$\sigma$ would be established . Note that, even in the modest case, observations at the discovery level could be achieved with next-generation detectors such as IceCube-Gen2.

Besides NGC 1068, the rest of the sources are unlikely to be identified in IceCube point source searches. However, stacking analyses would have sufficient sensitivities for identifying neutrino emission from the bright sources in the next few years in IceCube. Identification at $\sim3\sigma$ is plausible even when the two sources in our list with ambiguities on the contribution of coronal emission (Cen A and NGC 1275) are removed from a stacking analysis. 

\begin{figure}[t]
    \centering
    \includegraphics[width=\columnwidth]{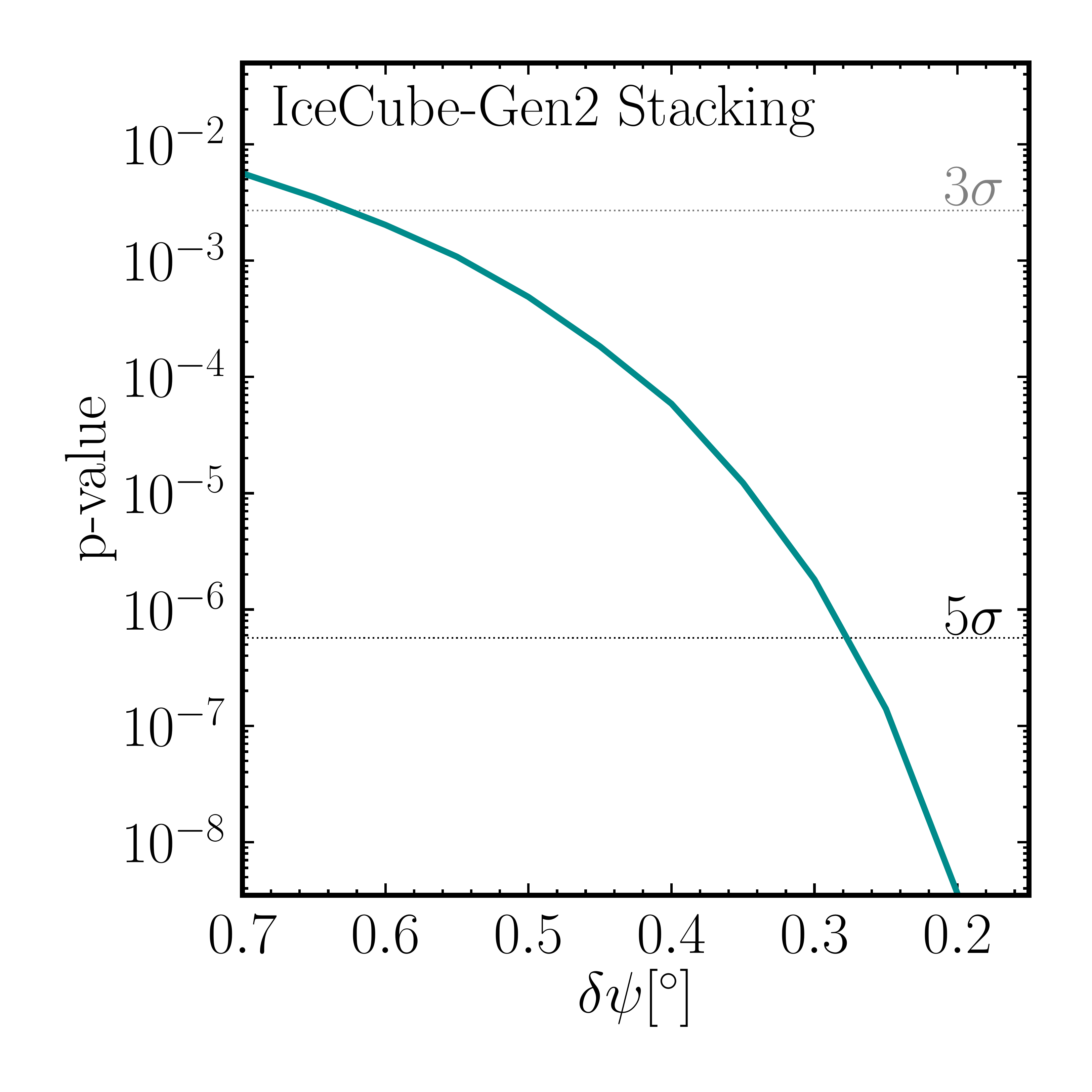}
    \caption{Dependence of the sensitivity of stacking analyses of bright nearby Seyfert galaxies in 5-year observation of IceCube-Gen2 for different angular resolutions in the medium-energy range. Here, we show the expected p-value for the stochastic acceleration scenario with the modest CR pressure.}
    \label{fig:pval_gen2_res}
\end{figure}

We further showed the prospects for identification of these sources in the stacking analysis in IceCube-Gen2, where we assume the modest CR pressure. While the increased effective area of the IceCube-Gen2 makes it possible to identify neutrino emission from these sources, the observational significance highly depends on the angular resolution of IceCube-Gen2 at $\sim10-30$ TeV energies. We showed the dependency of the prospects for observing neutrino emission from bright Seyferts in our list on the angular resolution of IceCube-Gen2 in Fig.~\ref{fig:pval_gen2_res}. 5$\sigma$ identification could be achieved with better than 0.4$^\circ$ angular uncertainty. 
This result further strengthens the importance of improving the angular resolution to identify the ``dominant'' origin of IceCube neutrinos \citep{Murase:2016gly}.

As the majority of the bright sources are in the Southern sky, the operation of KM3NeT in the near future will make it possible to identify neutrino emission from Cen A and Circinus Galaxy, even with the modest CR pressure case. Again, the signal from the rest of the sources is not strong enough to yield a significant excess. However, even the modest scenario should be identified within the five years of its operation. We should also note that in addition to KM3NeT, commissioning of the Baikal underwater neutrino telescope (NT-200)~\citep{Belolaptikov:1997ry,Aynutdinov:2006ca} and the Pacific-Ocean Neutrino Experiment (P-ONE) \citep{Agostini:2020aar} will boost the coverage for the sources in the Southern hemisphere. This would enhance the likelihood for identification of neutrino emission from sources considered in this study.

It is worth mentioning that the recent progress in enhancing IceCube's sensitivity for sources in the Southern sky can increase the likelihood for observation of the neutrino emission from Seyfert galaxies in the near future \citep{Mancina:2019hsp}. The technique enhances the exposure of IceCube in the medium-energy range in the 10 TeV range, which creates a unique opportunity to examine the stochastic acceleration scenario. 

In this study, we considered two acceleration mechanisms in the disk-corona model of AGN: stochastic acceleration and magnetic reconnection. They phenomenologically represent a hard CR spectrum and a broad power-law spectrum, respectively. 
One of the major differences between the spectra of high-energy neutrinos in these two scenarios is the level of the flux at energies above 100 TeV. Compared to the stochastic acceleration scenario with the high CR pressure, the flux normalization in the magnetic reconnection scenario is, in general, lower. This gives the stochastic acceleration scenario a better likelihood of observation in point source studies. However, in a stacking search, the accumulation of high-energy events from all sources leads to a stronger separation of signal and background events. Therefore, the magnetic reconnection scenario yields better p-values. The better likelihood also means a stronger constraint in the absence of evidence for a neutrino spectrum that extends to higher energies from these sources. { At the moment, the excess of events found in the direction of NGC 1068 is not statistically significant enough to constrain any of the scenarios discussed here. However, the confirmation of the steep spectrum reported by the IceCube Collaboration in the future will put constraints on particle acceleration mechanisms with a hard CR injection spectra.} For example, shock acceleration in the Bohm limit may be strongly constrained.  

The coronal neutrino emission model for the nearby, bright sources considered in this study is expected to be identified by current and future neutrino detectors. The magnitude of the diffuse neutrino flux observed in IceCube \citep{Aartsen:2020aqd} demonstrated that hadronuclear or photohadronic interactions play significant roles in the high-energy universe. The recent developments in the search for the origin of high-energy neutrinos have revealed the first sign of anisotropies in the flux of high-energy neutrinos \citep{Aartsen:2019fau}. At very high energies, events beyond 100 TeV have facilitated time-dependent searches and follow-ups, delivering coincidences with flaring blazars and tidal disruption events \citep{IceCube:2018dnn, Stein:2020xhk}. In the meantime, the higher level of the diffuse neutrino flux at medium energies implies the possibility of finding sources with a high-level of neutrino flux. Identifying their origin will bestow a unique opportunity to probe extraordinary environments in the non-thermal universe and particle acceleration in the dense environments that cannot be readily probed by electromagnetic observations.\\

\section*{Acknowledgements}
We would like to thank Chad Finley, Francis Halzen, Maria Petropoulou, and Ibrahim Safa for useful comments and discussion. 
A.K. acknowledges the support from the IGC through the IGC postdoctoral fellowship. 
The work of K.M. is supported by NSF Grant No.~AST-1908689, and KAKENHI No.~20H01901 and No.~20H05852. 
S.S.K. acknowledges JSPS Research Fellowship and KAKENHI Grant No.~19J00198.

\bibliography{bibfile}

\begin{thebibliography}{}
\expandafter\ifx\csname natexlab\endcsname\relax\def\natexlab#1{#1}\fi
\providecommand{\url}[1]{\href{#1}{#1}}

\bibitem[{Sen(2015)}]{Senno:2015tra}
 2015, Astrophys. J., 806, 24

\bibitem[{Aartsen {et~al.}(2013{\natexlab{a}})}]{Aartsen:2013bka}
Aartsen, M., {et~al.} 2013{\natexlab{a}}, Phys.Rev.Lett., 111, 021103

\bibitem[{Aartsen {et~al.}(2018)}]{IceCube:2018dnn}
---. 2018, Science, 361, eaat1378

\bibitem[{Aartsen {et~al.}(2020{\natexlab{a}})}]{Aartsen:2020aqd}
---. 2020{\natexlab{a}}, arXiv:2001.09520

\bibitem[{Aartsen {et~al.}(2020{\natexlab{b}})}]{Aartsen:2019fau}
---. 2020{\natexlab{b}}, Phys. Rev. Lett., 124, 051103

\bibitem[{Aartsen {et~al.}(2013{\natexlab{b}})}]{Aartsen:2013jdh}
Aartsen, M.~G., {et~al.} 2013{\natexlab{b}}, Science, 342, 1242856

\bibitem[{Aartsen {et~al.}(2015{\natexlab{a}})}]{Aartsen:2014muf}
---. 2015{\natexlab{a}}, Phys. Rev., D91, 022001

\bibitem[{Aartsen {et~al.}(2015{\natexlab{b}})}]{Aartsen:2015knd}
---. 2015{\natexlab{b}}, Astrophys.\ J., 809, 98

\bibitem[{Aartsen {et~al.}(2017)}]{Aartsen:2016oji}
---. 2017, Astrophys. J., 835, 151

\bibitem[{Abdo {et~al.}(2010)}]{Falcone:2010fk}
Abdo, A.~A., {et~al.} 2010, Astrophys. J., 719, 1433

\bibitem[{Ackermann {et~al.}(2012)Ackermann, Ajello, Allafort, Baldini, Ballet,
  Bastieri, Bechtol, Bellazzini, Berenji, Bloom, \& et~al.}]{Ackermann_2012}
Ackermann, M., Ajello, M., Allafort, A., {et~al.} 2012, The Astrophysical
  Journal, 755, 164.
\newblock \url{http://dx.doi.org/10.1088/0004-637X/755/2/164}

\bibitem[{Ackermann {et~al.}(2015)}]{Ackermann:2014usa}
Ackermann, M., {et~al.} 2015, Astrophys. J., 799, 86

\bibitem[{Adrian-Martinez {et~al.}(2016)}]{Adrian-Martinez:2016fdl}
Adrian-Martinez, S., {et~al.} 2016, J. Phys., G43, 084001

\bibitem[{Agostini {et~al.}(2020)}]{Agostini:2020aar}
Agostini, M., {et~al.} 2020, arXiv:2005.09493

\bibitem[{Albert {et~al.}(2020)}]{Aartsen:2020xpf}
Albert, A., {et~al.} 2020, Astrophys. J., 892, 92

\bibitem[{Alexandreas {et~al.}(1993)}]{Alexandreas:1992ek}
Alexandreas, D., {et~al.} 1993, Nucl. Instrum. Meth. A, 328, 570

\bibitem[{ATLAS(2011)}]{ATLAS:2011tau}
ATLAS, CMS, L. H. C.~G. 2011

\bibitem[{Aynutdinov {et~al.}(2006)}]{Aynutdinov:2006ca}
Aynutdinov, V.~M., {et~al.} 2006, Phys. Atom. Nucl., 69, 1914, [,449(2006)]

\bibitem[{Balbus \& Hawley(1991)}]{Balbus:1991ay}
Balbus, S.~A., \& Hawley, J.~F. 1991, Astrophys. J., 376, 214

\bibitem[{Ball {et~al.}(2018{\natexlab{a}})Ball, Sironi, \& Azel}]{Ball_2018}
Ball, D., Sironi, L., \& Azel, F. 2018{\natexlab{a}}, The Astrophysical
  Journal, 862, 80.
\newblock \url{https://doi.org/10.3847/1538-4357/aac820}

\bibitem[{Ball {et~al.}(2018{\natexlab{b}})Ball, Sironi, \&
  \"Ozel}]{Ball:2018icx}
Ball, D., Sironi, L., \& \"Ozel, F. 2018{\natexlab{b}}, Astrophys. J., 862, 80

\bibitem[{Bechtol {et~al.}(2017)Bechtol, Ahlers, Di~Mauro, Ajello, \&
  Vandenbroucke}]{Bechtol:2015uqb}
Bechtol, K., Ahlers, M., Di~Mauro, M., Ajello, M., \& Vandenbroucke, J. 2017,
  Astrophys. J., 836, 47

\bibitem[{{Becker} {et~al.}(2011){Becker}, {Das}, \& {Le}}]{bdl11}
{Becker}, P.~A., {Das}, S., \& {Le}, T. 2011, \apj, 743, 47

\bibitem[{{Begelman} {et~al.}(1990){Begelman}, {Rudak}, \& {Sikora}}]{brs90}
{Begelman}, M.~C., {Rudak}, B., \& {Sikora}, M. 1990, \apj, 362, 38

\bibitem[{Belolaptikov {et~al.}(1997)}]{Belolaptikov:1997ry}
Belolaptikov, I.~A., {et~al.} 1997, Astropart. Phys., 7, 263

\bibitem[{{Berezinskii} \& {Ginzburg}(1981)}]{1981MNRAS.194....3B}
{Berezinskii}, V.~S., \& {Ginzburg}, V.~L. 1981, \mnras, 194, 3

\bibitem[{Blackman \& Pessah(2009)}]{Blackman:2009fi}
Blackman, E.~G., \& Pessah, M.~E. 2009, Astrophys. J. Lett., 704, L113

\bibitem[{Capanema {et~al.}(2020{\natexlab{a}})Capanema, Esmaili, \&
  Murase}]{Capanema:2020rjj}
Capanema, A., Esmaili, A., \& Murase, K. 2020{\natexlab{a}}, Phys. Rev. D, 101,
  103012

\bibitem[{Capanema {et~al.}(2020{\natexlab{b}})Capanema, Esmaili, \&
  Serpico}]{Capanema:2020oet}
Capanema, A., Esmaili, A., \& Serpico, P.~D. 2020{\natexlab{b}},
  arXiv:2007.07911

\bibitem[{Chattopadhyay \& Kumar(2016)}]{Chattopadhyay:2016kcz}
Chattopadhyay, I., \& Kumar, R. 2016, Mon. Not. Roy. Astron. Soc., 459, 3792

\bibitem[{{Comisso} \& {Sironi}(2018)}]{2018PhRvL.121y5101C}
{Comisso}, L., \& {Sironi}, L. 2018, \prl, 121, 255101

\bibitem[{Comisso \& Sironi(2019)}]{Comisso:2019frj}
Comisso, L., \& Sironi, L. 2019, Astrophys. J., 886, 122

\bibitem[{{de Gouveia dal Pino} \& {Lazarian}(2005)}]{2005A&A...441..845D}
{de Gouveia dal Pino}, E.~M., \& {Lazarian}, A. 2005, \aap, 441, 845

\bibitem[{{del Valle} {et~al.}(2016){del Valle}, {de Gouveia Dal Pino}, \&
  {Kowal}}]{2016MNRAS.463.4331D}
{del Valle}, M.~V., {de Gouveia Dal Pino}, E.~M., \& {Kowal}, G. 2016, \mnras,
  463, 4331

\bibitem[{{Drake} {et~al.}(2006){Drake}, {Swisdak}, {Che}, \&
  {Shay}}]{2006Natur.443..553D}
{Drake}, J.~F., {Swisdak}, M., {Che}, H., \& {Shay}, M.~A. 2006, \nat, 443, 553

\bibitem[{Eichmann \& Becker~Tjus(2016)}]{Eichmann:2015ama}
Eichmann, B., \& Becker~Tjus, J. 2016, Astrophys. J., 821, 87

\bibitem[{Fang \& Murase(2018)}]{Fang:2017zjf}
Fang, K., \& Murase, K. 2018, Nature Phys., 14, 396

\bibitem[{{Giannios}(2010)}]{2010MNRAS.408L..46G}
{Giannios}, D. 2010, \mnras, 408, L46

\bibitem[{Gonzalez-Garcia {et~al.}(2014)Gonzalez-Garcia, Halzen, \&
  Niro}]{Gonzalez-Garcia:2013iha}
Gonzalez-Garcia, M., Halzen, F., \& Niro, V. 2014, Astropart.Phys., 57-58, 39

\bibitem[{{Guo} {et~al.}(2016){Guo}, {Li}, {Li}, {Daughton}, {Zhang},
  {Lloyd-Ronning}, {Liu}, {Zhang}, \& {Deng}}]{2016ApJ...818L...9G}
{Guo}, F., {Li}, X., {Li}, H., {et~al.} 2016, \apjl, 818, L9

\bibitem[{Guo {et~al.}(2016)Guo, Li, Li, Daughton, Zhang, Lloyd-Ronning, Liu,
  Zhang, \& Deng}]{Guo_2016}
Guo, F., Li, X., Li, H., {et~al.} 2016, The Astrophysical Journal, 818, L9.
\newblock \url{https://doi.org/10.3847/2041-8205/818/1/l9}

\bibitem[{Guti\'errez {et~al.}(2021)Guti\'errez, Vieyro, \&
  Romero}]{Gutierrez:2021vnk}
Guti\'errez, E.~M., Vieyro, F.~L., \& Romero, G.~E. 2021, Astron. Astrophys.,
  649, A87

\bibitem[{Haardt \& Maraschi(1991)}]{Haardt:1991tp}
Haardt, F., \& Maraschi, L. M.~U. 1991, Astrophys. J. Lett., 380, L51

\bibitem[{Halzen {et~al.}(2016)Halzen, Kheirandish, \& Niro}]{Halzen:2016seh}
Halzen, F., Kheirandish, A., \& Niro, V. 2016, arXiv:1609.03072

\bibitem[{Harris {et~al.}(2010)Harris, Rejkuba, \& Harris}]{Harris:2009wj}
Harris, G.~L., Rejkuba, M., \& Harris, W.~E. 2010, Publ. Astron. Soc. Austral.,
  27, 457

\bibitem[{Honda {et~al.}(2007)Honda, Kajita, Kasahara, Midorikawa, \&
  Sanuki}]{Honda:2006qj}
Honda, M., Kajita, T., Kasahara, K., Midorikawa, S., \& Sanuki, T. 2007,
  Phys.Rev., D75, 043006

\bibitem[{Hoshino(2012)}]{PhysRevLett.108.135003}
Hoshino, M. 2012, Phys. Rev. Lett., 108, 135003.
\newblock \url{https://link.aps.org/doi/10.1103/PhysRevLett.108.135003}

\bibitem[{Hoshino(2013)}]{Hoshino:2013pza}
---. 2013, Astrophys. J., 773, 118

\bibitem[{{Hoshino}(2015)}]{hos15}
{Hoshino}, M. 2015, Physical Review Letters, 114, 061101

\bibitem[{Hoshino(2015)}]{PhysRevLett.114.061101}
Hoshino, M. 2015, Phys. Rev. Lett., 114, 061101.
\newblock \url{https://link.aps.org/doi/10.1103/PhysRevLett.114.061101}

\bibitem[{{Hoshino} \& {Lyubarsky}(2012)}]{HL12}
{Hoshino}, M., \& {Lyubarsky}, Y. 2012, \ssr, 173, 521

\bibitem[{Inoue {et~al.}(2020)Inoue, Khangulyan, \& Doi}]{Inoue:2019yfs}
Inoue, Y., Khangulyan, D., \& Doi, A. 2020, Astrophys. J. Lett., 891, L33

\bibitem[{{Inoue} {et~al.}(2019){Inoue}, {Khangulyan}, {Inoue}, \&
  {Doi}}]{2019ApJ...880...40I}
{Inoue}, Y., {Khangulyan}, D., {Inoue}, S., \& {Doi}, A. 2019, \apj, 880, 40

\bibitem[{Io \& Suzuki(2014)}]{Io:2013gja}
Io, Y., \& Suzuki, T.~K. 2014, Astrophys. J., 780, 46

\bibitem[{Janssen {et~al.}(2015)Janssen, Bruderer, Sturm, Contursi, Davies,
  Hailey-Dunsheath, Poglitsch, Genzel, Graci{\'{a}}-Carpio, Lutz, Tacconi,
  Fischer, Gonz{\'{a}}lez-Alfonso, Sternberg, Veilleux, Verma, \&
  Burtscher}]{Janssen_2015}
Janssen, A.~W., Bruderer, S., Sturm, E., {et~al.} 2015, The Astrophysical
  Journal, 811, 74.
\newblock \url{https://doi.org/10.1088/0004-637x/811/2/74}

\bibitem[{Jiang {et~al.}(2019)Jiang, Blaes, Stone, \& Davis}]{Jiang:2019bxn}
Jiang, Y.-F., Blaes, O., Stone, J., \& Davis, S.~W. 2019, arXiv:1904.01674

\bibitem[{{Jiang} {et~al.}(2019){Jiang}, {Blaes}, {Stone}, \&
  {Davis}}]{2019ApJ...885..144J}
{Jiang}, Y.-F., {Blaes}, O., {Stone}, J.~M., \& {Davis}, S.~W. 2019, \apj, 885,
  144

\bibitem[{Jiang {et~al.}(2014)Jiang, Stone, \& Davis}]{Jiang:2014wga}
Jiang, Y.-F., Stone, J.~M., \& Davis, S.~W. 2014, Astrophys. J., 784, 169

\bibitem[{{Kalashev} {et~al.}(2015){Kalashev}, {Semikoz}, \&
  {Tkachev}}]{2015JETP..120..541K}
{Kalashev}, O., {Semikoz}, D., \& {Tkachev}, I. 2015, Soviet Journal of
  Experimental and Theoretical Physics, 120, 541

\bibitem[{{Karachentsev} {et~al.}(2013){Karachentsev}, {Makarov}, \&
  {Kaisina}}]{2013AJ....145..101K}
{Karachentsev}, I.~D., {Makarov}, D.~I., \& {Kaisina}, E.~I. 2013, \aj, 145,
  101

\bibitem[{{Kimura} {et~al.}(2019{\natexlab{a}}){Kimura}, {Murase}, \&
  {M{\'e}sz{\'a}ros}}]{2019PhRvD.100h3014K}
{Kimura}, S.~S., {Murase}, K., \& {M{\'e}sz{\'a}ros}, P. 2019{\natexlab{a}},
  \prd, 100, 083014

\bibitem[{{Kimura} {et~al.}(2016){Kimura}, {Toma}, {Suzuki}, \&
  {Inutsuka}}]{KTS16a}
{Kimura}, S.~S., {Toma}, K., {Suzuki}, T.~K., \& {Inutsuka}, S.-i. 2016, \apj,
  822, 88

\bibitem[{{Kimura} {et~al.}(2014){Kimura}, {Toma}, \& {Takahara}}]{ktt14}
{Kimura}, S.~S., {Toma}, K., \& {Takahara}, F. 2014, \apj, 791, 100

\bibitem[{{Kimura} {et~al.}(2019{\natexlab{b}}){Kimura}, {Tomida}, \&
  {Murase}}]{2019MNRAS.485..163K}
{Kimura}, S.~S., {Tomida}, K., \& {Murase}, K. 2019{\natexlab{b}}, \mnras, 485,
  163

\bibitem[{Kotera {et~al.}(2009)Kotera, Allard, Murase, Aoi, Dubois,
  {et~al.}}]{Kotera:2009ms}
Kotera, K., Allard, D., Murase, K., {et~al.} 2009, Astrophys.J., 707, 370

\bibitem[{Lamastra {et~al.}(2019)Lamastra, Tavecchio, Romano, Landoni, \&
  Vercellone}]{Lamastra:2019zss}
Lamastra, A., Tavecchio, F., Romano, P., Landoni, M., \& Vercellone, S. 2019,
  Astropart. Phys., 112, 16

\bibitem[{Lemoine \& Malkov(2020)}]{Lemoine:2020bsk}
Lemoine, M., \& Malkov, M.~A. 2020, Mon. Not. Roy. Astron. Soc., 499, 4972

\bibitem[{Levinson \& Rieger(2011)}]{Levinson:2010fc}
Levinson, A., \& Rieger, F. 2011, Astrophys. J., 730, 123

\bibitem[{{Li} {et~al.}(2021){Li}, {Guo}, \& {Liu}}]{2021PhPl...28e2905L}
{Li}, X., {Guo}, F., \& {Liu}, Y.-H. 2021, Physics of Plasmas, 28, 052905

\bibitem[{Liu {et~al.}(2002{\natexlab{a}})Liu, Mineshige, Meyer,
  Meyer-Hofmeister, \& Kawaguchi}]{Liu:2002ts}
Liu, B., Mineshige, S., Meyer, F., Meyer-Hofmeister, E., \& Kawaguchi, T.
  2002{\natexlab{a}}, Astrophys. J., 575, 117

\bibitem[{Liu {et~al.}(2002{\natexlab{b}})Liu, Mineshige, \&
  Shibata}]{Liu:2002ps}
Liu, B., Mineshige, S., \& Shibata, K. 2002{\natexlab{b}}, Astrophys. J. Lett.,
  572, L173

\bibitem[{Liu {et~al.}(2018)Liu, Murase, Inoue, Ge, \& Wang}]{Liu:2017bjr}
Liu, R.-Y., Murase, K., Inoue, S., Ge, C., \& Wang, X.-Y. 2018, Astrophys. J.,
  858, 9

\bibitem[{Loeb \& Waxman(2006)}]{Loeb:2006tw}
Loeb, A., \& Waxman, E. 2006, JCAP, 0605, 003

\bibitem[{{Lynn} {et~al.}(2014){Lynn}, {Quataert}, {Chandran}, \&
  {Parrish}}]{lyn+14}
{Lynn}, J.~W., {Quataert}, E., {Chandran}, B.~D.~G., \& {Parrish}, I.~J. 2014,
  \apj, 791, 71

\bibitem[{{Machida} \& {Matsumoto}(2003)}]{MM03a}
{Machida}, M., \& {Matsumoto}, R. 2003, \apj, 585, 429

\bibitem[{{Maisack} {et~al.}(1993){Maisack}, {Johnson}, {Kinzer}, {Strickman},
  {Kurfess}, {Cameron}, {Jung}, {Grabelsky}, {Purcell}, \&
  {Ulmer}}]{1993ApJ...407L..61M}
{Maisack}, M., {Johnson}, W.~N., {Kinzer}, R.~L., {et~al.} 1993, \apjl, 407,
  L61

\bibitem[{Mancina \& Silva(2020)}]{Mancina:2019hsp}
Mancina, S., \& Silva, M. 2020, PoS, ICRC2019, 954

\bibitem[{Marinucci {et~al.}(2016)}]{Marinucci:2015fqo}
Marinucci, A., {et~al.} 2016, Mon. Not. Roy. Astron. Soc., 456, L94

\bibitem[{Medina-Torrejon {et~al.}(2021)Medina-Torrejon, de~Gouveia Dal~Pino,
  Kadowaki, Kowal, Singh, \& Mizuno}]{Medina-Torrejon:2020fae}
Medina-Torrejon, T.~E., de~Gouveia Dal~Pino, E.~M., Kadowaki, L. H.~S.,
  {et~al.} 2021, Astrophys. J., 908, 193

\bibitem[{Merloni \& Fabian(2001)}]{Merloni:2000gs}
Merloni, A., \& Fabian, A. 2001, Mon. Not. Roy. Astron. Soc., 321, 549

\bibitem[{Miller \& Stone(2000)}]{Miller:1999ix}
Miller, K., \& Stone, J. 2000, Astrophys. J., 534, 398

\bibitem[{Murase {et~al.}(2013)Murase, Ahlers, \& Lacki}]{Murase:2013rfa}
Murase, K., Ahlers, M., \& Lacki, B.~C. 2013, Phys.Rev., D88, 121301

\bibitem[{Murase {et~al.}(2016)Murase, Guetta, \& Ahlers}]{Murase:2015xka}
Murase, K., Guetta, D., \& Ahlers, M. 2016, Phys. Rev. Lett., 116, 071101

\bibitem[{Murase {et~al.}(2008)Murase, Inoue, \& Nagataki}]{Murase:2008yt}
Murase, K., Inoue, S., \& Nagataki, S. 2008, Astrophys.J., 689, L105

\bibitem[{Murase {et~al.}(2020)Murase, Kimura, \& Meszaros}]{Murase:2019vdl}
Murase, K., Kimura, S.~S., \& Meszaros, P. 2020, Phys. Rev. Lett., 125, 011101

\bibitem[{Murase \& Waxman(2016)}]{Murase:2016gly}
Murase, K., \& Waxman, E. 2016, Phys. Rev., D94, 103006

\bibitem[{{Narayan} {et~al.}(2012){Narayan}, {S{\"a}dowski}, {Penna}, \&
  {Kulkarni}}]{2012MNRAS.426.3241N}
{Narayan}, R., {S{\"a}dowski}, A., {Penna}, R.~F., \& {Kulkarni}, A.~K. 2012,
  \mnras, 426, 3241

\bibitem[{{Ohsuga} \& {Mineshige}(2011)}]{OM11a}
{Ohsuga}, K., \& {Mineshige}, S. 2011, \apj, 736, 2

\bibitem[{Peretti {et~al.}(2020)Peretti, Blasi, Aharonian, Morlino, \&
  Cristofari}]{Peretti:2019vsj}
Peretti, E., Blasi, P., Aharonian, F., Morlino, G., \& Cristofari, P. 2020,
  Mon. Not. Roy. Astron. Soc., 493, 5880

\bibitem[{Petropoulou {et~al.}(2019)Petropoulou, Sironi, Spitkovsky, \&
  Giannios}]{Petropoulou:2019bse}
Petropoulou, M., Sironi, L., Spitkovsky, A., \& Giannios, D. 2019,
  arXiv:1906.03297

\bibitem[{Pisokas {et~al.}(2018)Pisokas, Vlahos, \& Isliker}]{Pisokas:2017zxx}
Pisokas, T., Vlahos, L., \& Isliker, H. 2018, Astrophys. J., 852, 64

\bibitem[{{Porth} {et~al.}(2019){Porth}, {Chatterjee}, {Narayan}, {Gammie},
  {Mizuno}, {Anninos}, {Baker}, {Bugli}, {Chan}, {Davelaar}, {Del Zanna},
  {Etienne}, {Fragile}, {Kelly}, {Liska}, {Markoff}, {McKinney}, {Mishra},
  {Noble}, {Olivares}, {Prather}, {Rezzolla}, {Ryan}, {Stone}, {Tomei},
  {White}, {Younsi}, {Akiyama}, {Alberdi}, {Alef}, {Asada}, {Azulay}, {Baczko},
  {Ball}, {Balokovi{\'c}}, {Barrett}, {Bintley}, {Blackburn}, {Boland},
  {Bouman}, {Bower}, {Bremer}, {Brinkerink}, {Brissenden}, {Britzen},
  {Broderick}, {Broguiere}, {Bronzwaer}, {Byun}, {Carlstrom}, {Chael},
  {Chatterjee}, {Chen}, {Chen}, {Cho}, {Christian}, {Conway}, {Cordes},
  {Geoffrey}, {Crew}, {Cui}, {De Laurentis}, {Deane}, {Dempsey}, {Desvignes},
  {Doeleman}, {Eatough}, {Falcke}, {Fish}, {Fomalont}, {Fraga-Encinas},
  {Freeman}, {Friberg}, {Fromm}, {G{\'o}mez}, {Galison}, {Garc{\'\i}a},
  {Gentaz}, {Georgiev}, {Goddi}, {Gold}, {Gu}, {Gurwell}, {Hada}, {Hecht},
  {Hesper}, {Ho}, {Ho}, {Honma}, {Huang}, {Huang}, {Hughes}, {Ikeda}, {Inoue},
  {Issaoun}, {James}, {Jannuzi}, {Janssen}, {Jeter}, {Jiang}, {Johnson},
  {Jorstad}, {Jung}, {Karami}, {Karuppusamy}, {Kawashima}, {Keating},
  {Kettenis}, {Kim}, {Kim}, {Kim}, {Kino}, {Koay}, {Patrick}, {Koch}, {Koyama},
  {Kramer}, {Kramer}, {Krichbaum}, {Kuo}, {Lauer}, {Lee}, {Li}, {Li},
  {Lindqvist}, {Liu}, {Liuzzo}, {Lo}, {Lobanov}, {Loinard}, {Lonsdale}, {Lu},
  {MacDonald}, {Mao}, {Marrone}, {Marscher}, {Mart{\'\i}-Vidal}, {Matsushita},
  {Matthews}, {Medeiros}, {Menten}, {Mizuno}, {Moran}, {Moriyama},
  {Moscibrodzka}, {M{\"u}ller}, {Nagai}, {Nagar}, {Nakamura}, {Narayanan},
  {Natarajan}, {Neri}, {Ni}, {Noutsos}, {Okino}, {Oyama}, {{\"O}zel},
  {Palumbo}, {Patel}, {Pen}, {Pesce}, {Pi{\'e}tu}, {Plambeck}, {PopStefanija},
  {Preciado-L{\'o}pez}, {Psaltis}, {Pu}, {Ramakrishnan}, {Rao}, {Rawlings},
  {Raymond}, {Ripperda}, {Roelofs}, {Rogers}, {Ros}, {Rose}, {Roshanineshat},
  {Rottmann}, {Roy}, {Ruszczyk}, {Rygl}, {S{\'a}nchez},
  {S{\'a}nchez-Arguelles}, {Sasada}, {Savolainen}, {Schloerb}, {Schuster},
  {Shao}, {Shen}, {Small}, {Sohn}, {SooHoo}, {Tazaki}, {Tiede}, {Tilanus},
  {Titus}, {Toma}, {Torne}, {Trent}, {Trippe}, {Tsuda}, {van Bemmel}, {van
  Langevelde}, {van Rossum}, {Wagner}, {Wardle}, {Weintroub}, {Wex}, {Wharton},
  {Wielgus}, {Wong}, {Wu}, {Young}, {Young}, {Yuan}, {Yuan}, {Zensus}, {Zhao},
  {Zhao}, {Zhu}, \& {Event Horizon Telescope
  Collaboration}}]{2019ApJS..243...26P}
{Porth}, O., {Chatterjee}, K., {Narayan}, R., {et~al.} 2019, \apjs, 243, 26

\bibitem[{Ricci {et~al.}(2017)}]{Ricci:2017dhj}
Ricci, C., {et~al.} 2017, Astrophys. J. Suppl., 233, 17

\bibitem[{{Ricci} {et~al.}(2018){Ricci}, {Ho}, {Fabian}, {Trakhtenbrot},
  {Koss}, {Ueda}, {Lohfink}, {Shimizu}, {Bauer}, {Mushotzky}, {Schawinski},
  {Paltani}, {Lamperti}, {Treister}, \& {Oh}}]{2018MNRAS.480.1819R}
{Ricci}, C., {Ho}, L.~C., {Fabian}, A.~C., {et~al.} 2018, \mnras, 480, 1819

\bibitem[{Sanuki {et~al.}(2007)Sanuki, Honda, Kajita, Kasahara, \&
  Midorikawa}]{Sanuki:2006yd}
Sanuki, T., Honda, M., Kajita, T., Kasahara, K., \& Midorikawa, S. 2007, Phys.
  Rev., D75, 043005

\bibitem[{Stecker {et~al.}(1991)Stecker, Done, Salamon, \&
  Sommers}]{Stecker:1991vm}
Stecker, F., Done, C., Salamon, M., \& Sommers, P. 1991, Phys.Rev.Lett., 66,
  2697

\bibitem[{Stecker(2005)}]{Stecker:2005hn}
Stecker, F.~W. 2005, Phys. Rev. D, 72, 107301

\bibitem[{Stecker(2013)}]{Stecker:2013fxa}
---. 2013, Phys. Rev. D, 88, 047301

\bibitem[{Stein {et~al.}(2020)}]{Stein:2020xhk}
Stein, R., {et~al.} 2020, arXiv:2005.05340

\bibitem[{{Stone} \& {Pringle}(2001)}]{SP01a}
{Stone}, J.~M., \& {Pringle}, J.~E. 2001, \mnras, 322, 461

\bibitem[{Tachibana {et~al.}(2015)Tachibana, Kawamuro, Ueda, Shidatsu, Arimoto,
  Yoshii, Yatsu, Saito, Pike, \& Kawai}]{Tachibana:2015ita}
Tachibana, Y., Kawamuro, T., Ueda, Y., {et~al.} 2015, arXiv:1504.03208

\bibitem[{{Tully}(1988)}]{1988ngc..book.....T}
{Tully}, R.~B. 1988, {Nearby galaxies catalog}

\bibitem[{{Tully} {et~al.}(2013){Tully}, {Courtois}, {Dolphin}, {Fisher},
  {H{\'e}raudeau}, {Jacobs}, {Karachentsev}, {Makarov}, {Makarova},
  {Mitronova}, {Rizzi}, {Shaya}, {Sorce}, \& {Wu}}]{2013AJ....146...86T}
{Tully}, R.~B., {Courtois}, H.~M., {Dolphin}, A.~E., {et~al.} 2013, \aj, 146,
  86

\bibitem[{Waxman \& Bahcall(1999)}]{Waxman:1998yy}
Waxman, E., \& Bahcall, J.~N. 1999, Phys.Rev., D59, 023002

\bibitem[{{Werner} {et~al.}(2018){Werner}, {Uzdensky}, {Begelman}, {Cerutti},
  \& {Nalewajko}}]{2018MNRAS.473.4840W}
{Werner}, G.~R., {Uzdensky}, D.~A., {Begelman}, M.~C., {Cerutti}, B., \&
  {Nalewajko}, K. 2018, \mnras, 473, 4840

\bibitem[{Wong {et~al.}(2020)Wong, Zhdankin, Uzdensky, Werner, \&
  Begelman}]{Wong:2019dog}
Wong, K., Zhdankin, V., Uzdensky, D.~A., Werner, G.~R., \& Begelman, M.~C.
  2020, Astrophys. J. Lett., 893, L7

\bibitem[{Xiao {et~al.}(2016)Xiao, M\'esz\'aros, Murase, \& Dai}]{Xiao:2016rvd}
Xiao, D., M\'esz\'aros, P., Murase, K., \& Dai, Z.-g. 2016, Astrophys. J., 826,
  133

\bibitem[{Yoast-Hull {et~al.}(2014)Yoast-Hull, III, Zweibel, \&
  Everett}]{Yoast-Hull:2013qfa}
Yoast-Hull, T.~M., III, J., Zweibel, E.~G., \& Everett, J.~E. 2014, Astrophys.
  J., 780, 137

\bibitem[{{Zdziarski}(1986)}]{1986ApJ...305...45Z}
{Zdziarski}, A.~A. 1986, \apj, 305, 45

\bibitem[{{Zdziarski} {et~al.}(2000){Zdziarski}, {Poutanen}, \&
  {Johnson}}]{2000ApJ...542..703Z}
{Zdziarski}, A.~A., {Poutanen}, J., \& {Johnson}, W.~N. 2000, \apj, 542, 703

\bibitem[{Zhang {et~al.}(2021)Zhang, Sironi, \& Giannios}]{Zhang:2021akj}
Zhang, H., Sironi, L., \& Giannios, D. 2021, arXiv:2105.00009

\bibitem[{Zhdankin {et~al.}(2019)Zhdankin, Uzdensky, Werner, \&
  Begelman}]{Zhdankin:2018lhq}
Zhdankin, V., Uzdensky, D.~A., Werner, G.~R., \& Begelman, M.~C. 2019, Phys.
  Rev. Lett., 122, 055101

\end{thebibliography}
\section{Appendix}

\subsection{Cosmic-ray luminosity}
\begin{figure}[h]
    \centering
    \includegraphics[width=0.5\columnwidth]{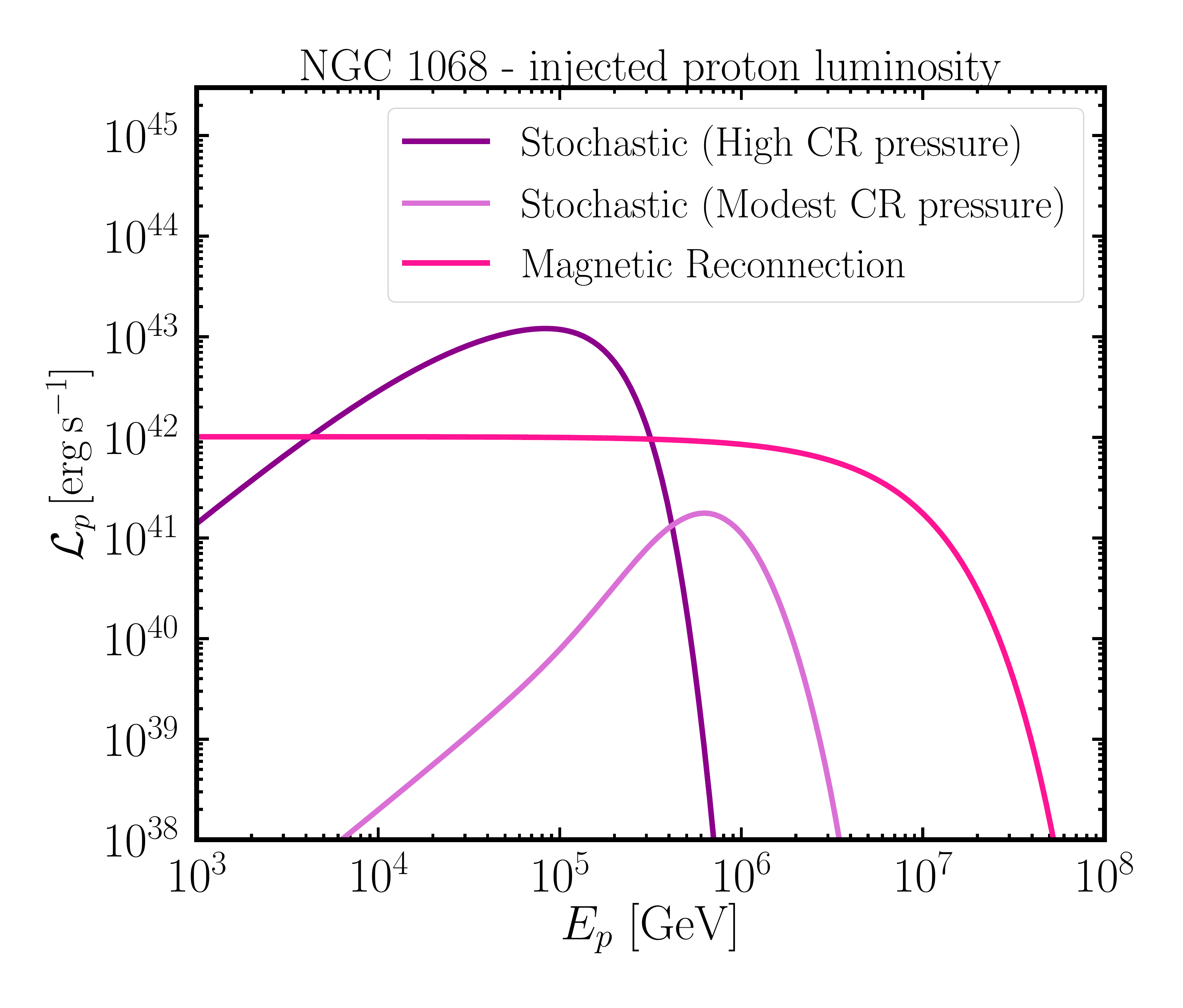}
    \caption{Differential CR proton luminosities for NGC 1068 ($L_X = 10^{43} \rm erg\,s^{-1}$) for the three different scenarios shown in Fig.~\ref{fig:ngc1068nuflux} and parameters in Table~\ref{tab:params}.}
    \label{fig:ngc1068CRLum}
\end{figure}

The differential CR luminosity for the three scenarios for particle acceleration in NGC 1068 are shown in Fig.~\ref{fig:ngc1068CRLum}. The CR luminosity is deduced from the measured intrinsic X-ray by incorporating the escape and cooling timescales, injection function for the CRs, and the volume of the coronal region. For more details, see Supplemental Material of \cite{Murase:2019vdl}. 

\subsection{Event distributions}
To calculate the number of background atmospheric neutrino events, we have integrated the atmospheric flux \cite{Honda:2006qj} and \cite{Sanuki:2006yd} over an opening angle of $\Omega=\pi \sigma^2$ around the direction of the source, where the angle $\sigma$ is the angular resolution of the detector. Given that the majority of the events in our models appear below $\sim$ 10 TeV, we set the angular resolution for the IceCube and KM3NeT to $0.7^\circ$ as specified in Fig.~2 of \cite{Aartsen:2016oji} and Fig.~22 of \cite{Adrian-Martinez:2016fdl}, respectively. 

\begin{figure*}[h]
    \centering
    \includegraphics[width=\textwidth]{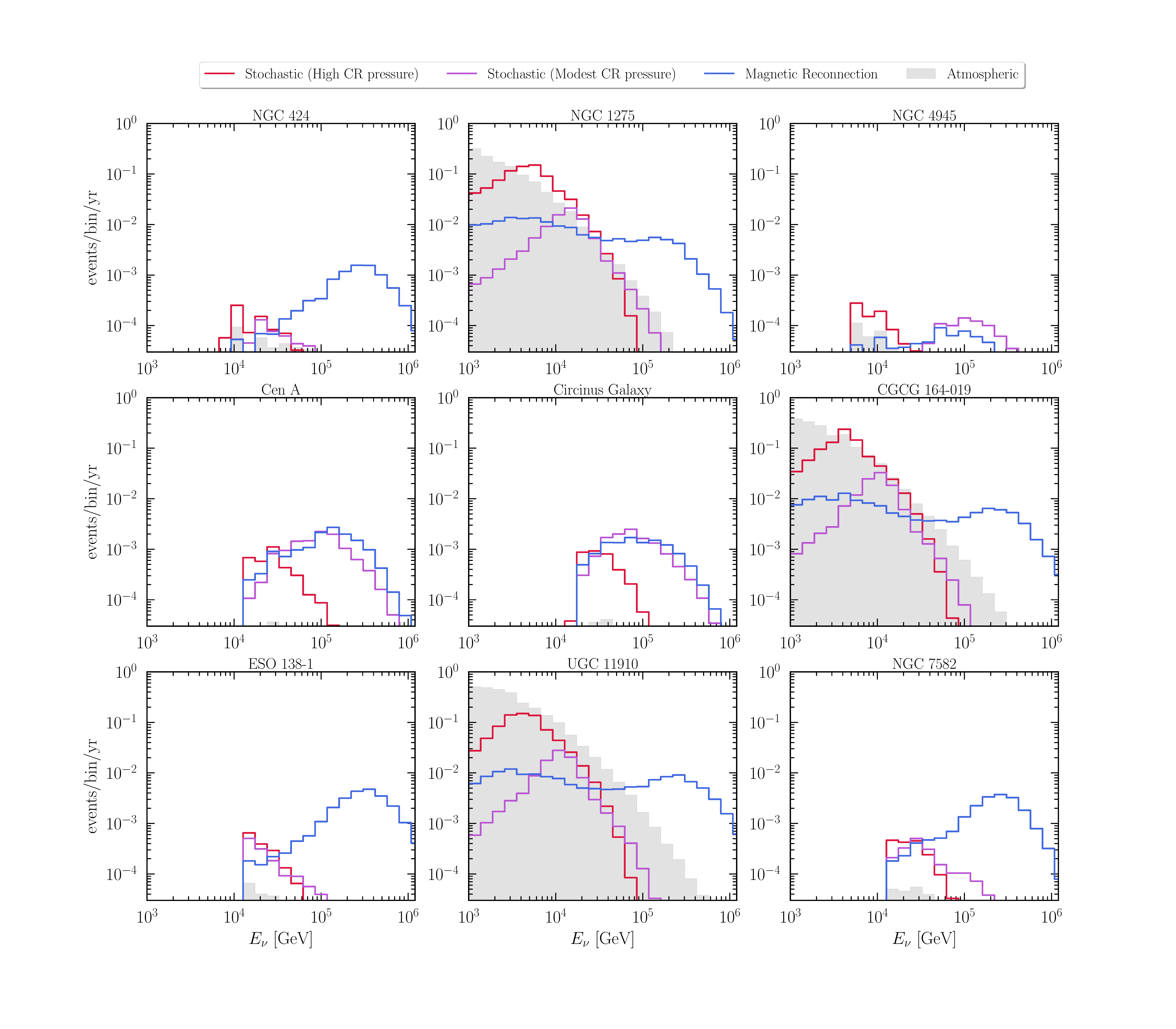}
    \caption{Expected event distribution in IceCube per year period for the three scenarios considered in this study. The expected background component from conventional atmospheric neutrino flux is shown for each source. For the sources located in the southern hemisphere, the event at low energies are suppressed due to the strict cuts in IceCube point source selection. The lines in each panel correspond to the fluxes presented in Fig.~\ref{fig:nuflux-all}}
    \label{fig:event_ic_all}
\end{figure*}

To calculate the number of cosmic neutrino events for the flux predicted in each scenario, we integrate the product of the neutrino spectrum and effective area over energy and time. For IceCube, we use the zenith dependent effective area for the point source analysis \citep{Aartsen:2016oji}. Similarly, we use this effective area, together with the atmospheric neutrino flux of \cite{Honda:2006qj} and \cite{Sanuki:2006yd} to estimate the number of background events in the direction of each source within the opening angle of $\Omega=\pi \sigma^2$. The signal and background event distributions for NGC 1068 is presented in Fig.~\ref{fig:ngc1068_ic_events} in the main text, and in Fig.~\ref{fig:event_ic_all} for the rest of bright Seyfert galaxies in Table~\ref{tab:sources}. Given that the majority of the sources in Fig.~\ref{fig:event_ic_all} are located in the Southern hemisphere, the corresponding event distribution is limited by the strong event selection cuts imposed to remove the atmospheric muon background penetrating the detector.

\begin{figure*}[h]
    \centering
    \includegraphics[width=\textwidth]{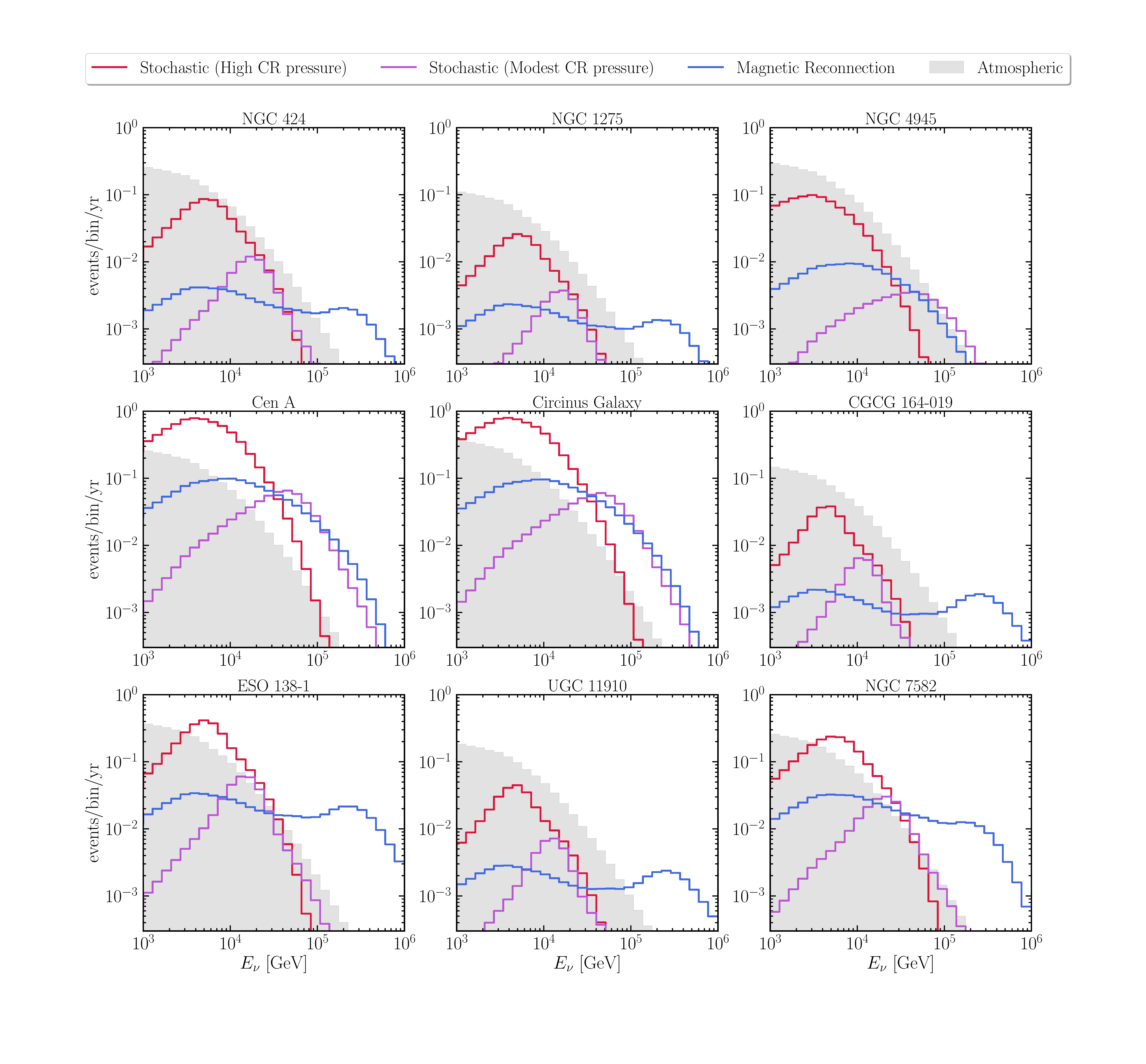}
    \caption{Signal and background events per energy in 1 year of KM3NeT for the 3 different schemes of neutrino emission considered in this study. The lines in each panel correspond to the fluxes presented in Fig.~\ref{fig:nuflux-all}.}
    \label{fig:events_all_km3net}
\end{figure*}

The event distribution for KM3NeT is presented in Fig.~\ref{fig:events_all_km3net}. We incorporate the average effective area for upgoing events in KM3NeT according to Fig.~19 of \cite{Adrian-Martinez:2016fdl} and apply the trigger efficiency for the points source searches. To take into account the fraction of the duration for which the source is below the horizon for KM3NeT, we rescale the number of events by the visibility of the sources, given by their declination as is defined in Fig.~37 of \cite{Adrian-Martinez:2016fdl}. 

\clearpage

\subsection{Estimation of the Statistical significance}
We estimate the statistical significance for observing the sources using the analytic expression introduced by~\cite{ATLAS:2011tau}. This method has been used extensively to evaluate the prospects for observation of Galactic sources of high-energy neutrinos, see e.g., \citet{Gonzalez-Garcia:2013iha,Halzen:2016seh}. In this method, the p-value of identifying signal events from a source is given by
\begin{equation}
p_{\rm value}=\frac{1}{2}\left[ 1-{\rm{erf}} 
\left( \sqrt{q_0^{obs}/2} \right) \right]\,,
\end{equation}
where $q_0^{obs}$ is defined as 
\begin{equation}
q_0^{obs} \equiv -2 \ln \mathcal{L}_{b,D}= 2 \sum_i \left( Y_{b,i} - N_{D,i} + N_{D,i} \ln \left( \frac{N_{D,i}}{Y_{b,i}}\right)\right)\, ,
\end{equation}
with $i$ running over the different energy bins. Here, $Y_{b,i}$ is the theoretical expectation for the background hypothesis, while $N_{D,i}$ is the estimated signal generated as the median of events Poisson-distributed around the signal plus background. The background expectation here is the number of atmospheric neutrinos in the solid angle set by the angular resolution. This solid angle corresponds roughly 72\% of the signal events from the source; see \cite{Alexandreas:1992ek} for a discussion. 

\end{document}